\documentclass{quantumview}
\usepackage[utf8]{inputenc}
\usepackage[english]{babel}
\usepackage[T1]{fontenc}

\usepackage[numbers,sort&compress]{natbib}
\usepackage{lipsum}
\usepackage{mdframed}
\usepackage{booktabs}
\usepackage{longtable}

\usepackage{pdflscape} 
\usepackage{tabularx} 
\usepackage{booktabs} 
\usepackage{amssymb} 
\usepackage{algorithm}
\usepackage{algorithmicx} 
\usepackage{algpseudocode}  
\usepackage{amsmath} 
\usepackage{multirow}

\begin{document}
	\title{Scalable Parallel Simulation of Quantum Circuits on CPU and GPU Systems}

    \author{Guolong Zhong}
    \affiliation{School of Software Engineering, University of Science and Technology of China, 1129 Huizhou Avenue, Hefei 230051, Anhui, China}
    
    \author{Yi Fan}
    \email{fanyi@ustc.edu.cn}
    \affiliation{Hefei National Research Center for Physical Sciences at the Microscale, University of Science and Technology of China, 96 Jinzhai Road, Hefei 230026, Anhui, China}
    
    \author{Zhenyu Li}
    \email{zyli@ustc.edu.cn}
    \affiliation{State Key Laboratory of Precision and Intelligent Chemistry, University of Science and Technology of China, 96 Jinzhai Road, Hefei 230026, Anhui, China}


    
    
    
    

    \begin{abstract}
         Quantum computing enables parallelism through superposition and entanglement and offers advantages over classical computing architectures. However, due to the limitations of current quantum hardware in the noisy intermediate-scale quantum (NISQ) era, classical simulation remains a critical tool for developing quantum algorithms. In this research, we present a comprehensive parallelization solution for the Q$^2$Chemistry software package, delivering significant performance improvements for the full-amplitude simulator on both CPU and GPU platforms. By incorporating batch-buffered overlap processing, dependency-aware gate contraction and staggered multi-gate parallelism, our optimizations significantly enhance the simulation speed compared to unoptimized baselines, demonstrating the effectiveness of hybrid-level parallelism in HPC systems. Benchmark results show that Q$^2$Chemistry consistently outperforms current state-of-the-art open-source simulators across various circuit types. These benchmarks highlight the capability of Q$^2$Chemistry to effectively handle large-scale quantum simulations with high efficiency and high portability.
    \end{abstract}

    
    \keywords{Quantum Computing, Quantum Simulation, High-performance Computing, Heterogeneous systems}

	\maketitle


	\clearpage

	\section{Introduction}
	The central challenge in quantum chemistry lies in solving the electronic structure problem through numerical solutions of the Schrödinger equation. Over the past century, researchers have developed several computational methods on supercomputers, including wavefunction-based approaches such as Hartree-Fock (HF), coupled-cluster (CC) theory, and configuration interaction (CI), as well as the widely adopted density functional theory (DFT) frameworks \cite{Hon64, Kohn65, Kohn96, CohMorYan12}. These methods, while highly successful, face an intrinsic trade-off between computational accuracy and efficiency. Fast, approximate methods such as orbital-free DFT and machine-learned neural network potentials allow large-scale simulations at the order of millions of atoms, yet often fails in strongly correlated systems. In contrast, high-accuracy methods such as full configuration interaction (FCI) exhibit exponential scaling with system size, limiting their applicability to only small active spaces (e.g., 24 electrons in 24 orbitals) even on top-tier supercomputers \cite{casscf24o24e}. A simple approach to push forward the boundary of quantum chemistry simulations is increasing the computational power of high-performance computing (HPC) clusters. However, such progression is encountering critical bottlenecks. The manufacturing of transistors has nearly reached physical limits due to quantum tunneling, while enlarging the size of supercomputer clusters suffers from increased energy consumption and thermal dissipation. These constraints motivate the development of fundamentally new paradigms for computational chemistry.

    Quantum computing which was first proposed by Manin and Feynman in the early 1980s offers an innovative approach\cite{Feynman1982,  shor1997, h2molecule2010, Arute2019QuantumSU}. By leveraging quantum superposition and entanglement, quantum computing is supposed to simulate many-body quantum systems more efficiently and quantum chemistry is a particularly promising application. Among the quantum algorithms developed for this domain, hybrid quantum-classical algorithms including variational quantum eigensolver (VQE) \cite{Tilly_VQE_2021, Cerezo2021, Magann_VQA_2021, Fedorov_VQERev_2022, BraKit02, McAEndAsp20, CaoRomOls19, Preskill_2018, GeoAshNor14, AspDutLov05, Wang08, PerMcCSha14, HemMaiRom18, NamChen20, SheZhaZha17, MalBabKiv16, Kandala2017, ColRamDah18, McCRomBab16, LanWhi10, RomBabMcC18, vqe-excited-vqd, Mcclean17qse, YungCas2014, farhi2014} has emerged with great application potential. VQE encodes molecular wavefunctions into quantum states, maps electronic Hamiltonians to sums of Pauli operators, and optimizes parameterized quantum circuits in an iterative quantum-classical loop. Physically motivated ans\"{a}tze such as unitary coupled-cluster (UCC) and hardware-efficient circuits enable chemically meaningful simulations.\cite{BarKucNog89, TauBar06, Preskill_2018, Kandala2017} However, they often demand enormous quantum resources which far beyond the capability of current NISQ hardwares. For example, a typical UCCSD simulation of C$_3$H$_6$ in the STO-3G basis requires 42 entangled qubits and over $6.6 \times 10^6$ CNOT gates. Such an experimental setting significantly exceeds the performance of any existing quantum devices in both qubit scale and gate fidelity. Classical quantum simulators thus play a critical role in this paradigm by enabling algorithm development, testing, and validation on a classical computer before deployment on physical devices.

    The Schr\"{o}dinger-style full-amplitude simulation is one of the most important quantum circuit simulation methods. Unlike tensor-network based methods such as matrix product state simulators\cite{Guo2023differentiable, Vidal2007MPS, Hastings2009MPS} which comprise accuracy for larger simulation scale, full-amplitude method gives accurate results for simulating quantum circuit evolution, performing resource estimation, and exploring circuit design for small- and medium-scale quantum algorithms on a classical computer. There has been considerable research interests in developing such quantum simulators, leading to the creation of several software packages tailored for different computational objectives\cite{ Qibo, Avila2014GPUaware, 2019QuantumCS, Gutierrez, Gutirrez2008ParallelQC, 0.5petabyte, massive-parallel, qhipster, zulehner2018, Wu_2019, qiskit-package, Zhang-2015, cirq, qulacs, Jones_QUEST_2019}. However, existing full-amplitude simulators face three major limitations:
    (1) Inefficient gate operation execution due to sequential processing of high-density single- and two-qubit gates which often occur in ansatzes such as UCCSD and HEA; (2) Limited exploitation of data locality in distributed systems through fragmented optimization strategies; (3) Suboptimal GPU utilization from underdeveloped multi-dimensional single instruction multiple thread (SIMT) parallelism, as listed in Table~\ref{tab:simulator_features}. These fundamental limitations collectively hinder simulations of complex quantum circuits required for practical chemistry applications. To address these challenges, we present a series of parallel performance optimization solutions for the quantum chemistry simulation software Q$^2$Chemistry. Our contributions include:
    \begin{enumerate}
    \item \textbf{Batch-Buffered Overlap Processing (BBOP):} A multi-buffering strategy that overlaps data transfers with computations to minimize communication overhead in distributed systems. This approach partitions quantum state amplitudes into smaller batches and employs non-blocking MPI communication to achieve pipelined execution. 
    
    \item \textbf{Staggered Multi-Gate Parallelism (SMGP):} A two-dimensional thread block strategy for GPU execution that maximizes memory throughput by staggering gate operations across independent quantum state segments. This is a specifically designed optimization for muti-dimensional thread layout on GPUs, which avoids inter-thread conflicts while fully exploiting GPU concurrency.
    
    \item \textbf{Dependency-Aware Gate Contraction (DAGC):} A greedy algorithm that merges independent gates based on control-target dependencies, reducing operation counts through directed acyclic graph (DAG) analysis. This method is particularly effective for contracting small blocks of adjacent gates, leading to reduced circuit depth and operation counts in circuits with densely packed single-qubit operations, such as in UCCSD or HEA. 
    
    \end{enumerate} 
    
    Our framework uniquely combines these key features and enables Q$^2$Chemistry\cite{q2chem} to outperform existing simulators across all circuit types. We validate our optimizations through quantum chemistry benchmarks on CPU-only clusters and GPU heterogeneous systems. Benchmark results show substantial efficiency improvements by incorporating a suite of CPU-adaptable and GPU-adaptable optimization strategies. For 30-qubit QAOA circuits, our framework achieves a 3.01× CPU-side speedup and an additional 2.66× speedup on GPU compared with baseline. Similarly, for VQE-HEA circuits with 30 qubits, the combination of optimizations delivers up to a 4.52× speedup on CPU and 3.57× acceleration on GPU. Furthermore, cross-simulator benchmarks confirm that our framework delivers competitive performance across various types of circuits, achieving substantial speedup over existing open-source software packages. These results highlight the effectiveness of our hardware-agnostic optimization strategies for full-amplitude quantum circuit simulations.
    
    \begin{landscape}
    \begin{table}[htbp]
    \centering
    \caption{Feature comparison of selected quantum simulators}
    \label{tab:simulator_features}
    \renewcommand{\arraystretch}{1.2}
    \begin{tabularx}{\linewidth}{l*{9}{>{\centering\arraybackslash}X}}
    \toprule
    \textbf{Simulator} & \textbf{Version} & \textbf{MPI} & \textbf{OpenMP} & \textbf{AVX} & \textbf{Single-GPU} & \textbf{Multi-GPU} & \textbf{Gate Fusion} & \textbf{Pipelined} & \textbf{NCCL} \\
    \midrule
    Qiskit       & 1.4.0  & $\times$ & $\checkmark$ & $\checkmark$ & $\checkmark$ & $\checkmark$ & $\checkmark$ & $\times$ & $\times$ \\
    Qulacs       & 0.6.11 & $\checkmark$ & $\checkmark$ & $\checkmark$ & $\checkmark$ & $\times$ & $\checkmark$ & $\times$ & $\times$ \\
    Yao          & 0.7.4  & $\times$ & $\checkmark$ & $\times$ & $\checkmark$ & $\times$ & $\checkmark$ & $\times$ & $\times$ \\
    ProjectQ     & 0.8.0  & $\times$ & $\checkmark$ & $\checkmark$ & $\times$ & $\times$ & $\checkmark$ & $\times$ & $\times$ \\
    MindQuantum  & 0.10.0 & $\times$ & $\checkmark$ & $\times$ & $\times$ & $\times$ & $\checkmark$ & $\times$ & $\times$ \\
    Qibo         & 0.2.11 & $\times$ & $\checkmark$ & $\times$ & $\checkmark$ & $\times$ & $\times$ & $\times$ & $\times$ \\
    QuEST        & 4.0.0  & $\checkmark$ & $\checkmark$ & $\checkmark$ & $\checkmark$ & $\checkmark$ & $\times$ & $\times$ & $\checkmark$ \\
    Qsim         & 0.21.0 & $\checkmark$ & $\checkmark$ & $\checkmark$ & $\checkmark$ & $\checkmark$ & $\times$ & $\times$ & $\times$ \\
    Pennylane    & 0.38.0 & $\times$ & $\checkmark$ & $\times$ & $\checkmark$ & $\times$ & $\checkmark$ & $\times$ & $\times$ \\
    Q$^2$Chemistry     & 1.0.0  & $\checkmark$ & $\checkmark$ & $\checkmark$ & $\checkmark$ & $\checkmark$ & $\checkmark$ & $\checkmark$ & $\checkmark$ \\
    \bottomrule
    \end{tabularx}
    \end{table}
    \end{landscape}

    \section{Full-amplitude Simulation Algorithm}\label{sec2}
    \subsection{Linear Algebra of Quantum computing}\label{subsec2.1}
    Quantum states, quantum circuits and quantum gates constitute the key elements of circuit-based quantum computing. The quantum state can be represented by a statevector in the Hilbert space (denoted as $|\psi \rangle$), which is a linear superposition of basis vectors $|0\rangle, |1\rangle, \dots , |2^n-1\rangle$: 
    \begin{equation}
    	\begin{aligned}
    		|\psi\rangle = \alpha_{0} |\cdots00\rangle + \alpha_{1} |\cdots01\rangle + \alpha_{2} |\cdots10\rangle + \alpha_{3} |\cdots11\rangle + \cdots 
    		= \sum_{0}^{2^n-1} \alpha_i |i \rangle  
    		= \begin{pmatrix}
    			\alpha_{0} \\
    			\alpha_{1} \\
    			\alpha_{2} \\
    			\alpha_{3} \\
    			\vdots
    		\end{pmatrix},
    	\end{aligned}
    \end{equation}
    where $\sum_{0}^{2^n-1} |\alpha_i| ^ 2 = 1$.
    
    A quantum circuit is composed of several quantum gates in time sequence. Similar to the logic gates in classical computing, quantum gates are symbols for performing operations on qubits. The process of quantum state evolution can be expressed as $|\Psi_{out} \rangle = U|\Psi_{in} \rangle$, where $U$ is a $2^n \times 2^n$ unitary matrix, and $n$ is the number of qubits. Taking the single-qubit gate which has a $2\times 2$ matrix form $U_1$ as an example and denoting the index of target qubit as $k$, its operation on the quantum state can be expressed as:
    \begin{equation}
    	\begin{aligned}
    		U_g |\psi \rangle &= \overbrace{I_2 \otimes \cdots \otimes I_2}^{n-k-1} \otimes U_1 \otimes \overbrace{I_2 \otimes \cdots \otimes I_2}^{k} |\Psi \rangle\\
    		&= I_2^{\otimes n-k-1} \otimes U_1 \otimes I_2^{\otimes k} |\Psi \rangle,
    	\end{aligned}
    \end{equation}
    where $I_2$ represents the $2\times 2$ identity matrix.
    
    \begin{figure}[ht]
    	\centering
    	\includegraphics[width=0.8\textwidth]{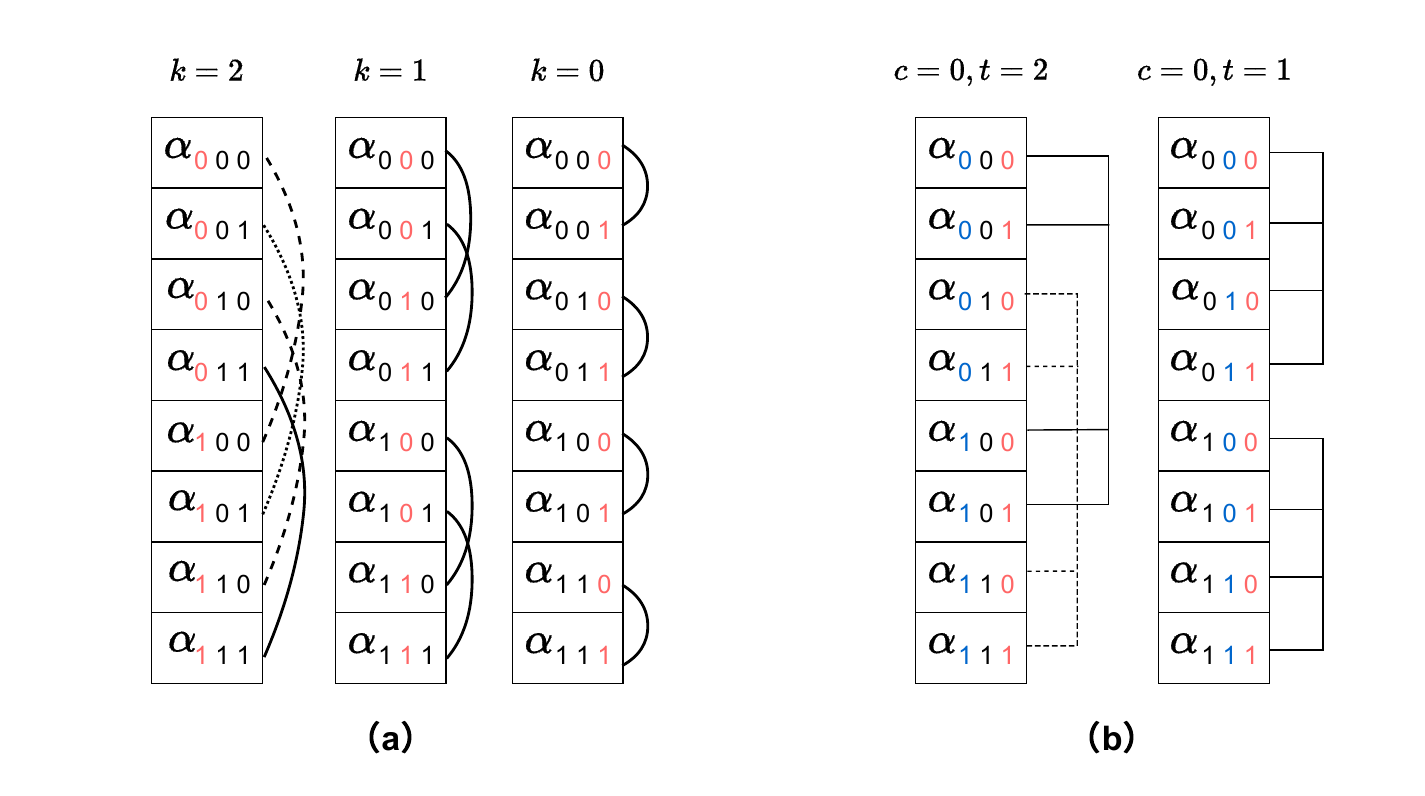}
        \caption{Illustration of quantum gate operations on a three-qubit system. (a) A single-qubit gate acting on the second qubit ($k$) of the quantum register. The operation affects pairs of amplitudes that differ only in the value of the target qubit ($k$). (b) A controlled gate with the control qubit set to $c$ and the target qubit to $t$. The gate is conditionally applied depending on the state of the control qubit, modifying only the relevant amplitudes in the quantum state vector.}\label{fig::gate_operation_1}
    \end{figure} 
    
    Instead of explicitly constructing the entire matrix $U_g$, $U_1$ can be applied on the statevector in a more efficient way\cite{qhipster}, as is illustrated in Figure \ref{fig::gate_operation_1}a where a three-qubit statevector is used as an example and the evolution of a single-qubit gate is converted to matrix-vector multiplications on pairs of quantum state amplitudes. A general formula is summarized as Equation \ref{eq3}. 
    \begin{equation}\label{eq3}
    	\begin{pmatrix}
    		\alpha'_{**0_k**} \\
    		\alpha'_{**1_k**}
    	\end{pmatrix}
    	= U_1
    	\begin{pmatrix}
    		\alpha_{**0_k**} \\
    		\alpha_{**1_k**}
    	\end{pmatrix}
    	= 
    	\begin{bmatrix}
    		u_{00} & u_{01} \\
    		u_{10} & u_{11}
    	\end{bmatrix}
    	\begin{pmatrix}
    		\alpha_{**0_k**} \\
    		\alpha_{**1_k**}
    	\end{pmatrix}
    	=
    	\begin{pmatrix}
    		u_{00} \alpha_{**0_k**} + u_{01} \alpha_{**1_k**} \\
    		u_{10} \alpha_{**0_k**} + u_{11} \alpha_{**1_k**}
    	\end{pmatrix}
    \end{equation}
    
    Similarly, the two-qubit controlled quantum gate CU also operates on groups of coefficients. Given the common matrix form of CU as:
    \begin{center}
    	$CU = |0\rangle \langle 0| \otimes I +|1\rangle \langle 1| \otimes U =  \begin{bmatrix}
    		1 & 0 & 0 & 0 \\
    		0 & 1 & 0 & 0 \\
    		0 & 0 & u_{00} & u_{01} \\
    		0 & 0 & u_{10} & u_{11}
    	\end{bmatrix}$,
    \end{center}
    the existence of the control qubit $c$ and target qubit $t$ makes every four amplitude as a group($|*0_t*0_c*\rangle, |*0_t*1_c*\rangle, |*1_t*0_c*\rangle, |*1_t*1_c*\rangle$), as shown in Figure \ref{fig::gate_operation_1}b. The CU gate will only operate and transform two of the amplitudes ($|*0_t*1_c*\rangle, |*1_t*1_c*\rangle$). Therefore, compared to a single-qubit gate evolution, CU only requires half of floating-point operations as well as data read/write amount. This process is summarized as Equation \ref{eq4}: 
    \begin{align}\label{eq4}
    	\begin{pmatrix}
    		\alpha'_{*0_t*1_c*} \\ \alpha'_{*1_t*1_c*}
    	\end{pmatrix}
    	&=
    	\begin{pmatrix}
    		u_{00} \alpha_{*0_t*1_c*} + u_{01} \alpha_{*1_t*1_c*}\\
    		u_{10} \alpha_{*0_t*1_c*} + u_{11} \alpha_{*1_t*1_c*}
    	\end{pmatrix}
    \end{align}
    
    \subsection{Implementation of a Full-amplitude Simulator}\label{subsec2.2}
    
    The key step in a full-amplitude simulation is to find all the amplitude pairs for matrix-vector calculations. For the target qubit $k$, we define its binary mask as $mask\_k = 2^k$. For one coefficient in the statevector, we apply bit-wise operations on its index $i$: $i \& mask\_k$. If the outcome is 0, then the indices of amplitude pairs can be calculated as $i \& mask\_k$ (for 0 state on qubit $k$) and $i + mask\_k$ (for 1 state on qubit $k$). In this way, we can find all pairs of coefficients by traversing all indices the binary form of which is equivalent to quantum state basis. Algorithm \ref{algo1} demonstrates the process of a single-qubit operation on a statevector. 
    Similarly, for CU gate operates on $c$-th qubit (the control qubit) and $t$-th qubit (the target qubit), we introduce $mask\_c = 2^c$ and $mask\_t = 2^t$. A similar procedure to perform statevector update can be easily obtained as Algorithm \ref{algo2}.
    
    \begin{algorithm}
    	\caption{Single-qubit Gate Operation}\label{algo1}
    	\begin{algorithmic}[1]
    		\Require State vector $\psi$, target qubit $k$, total qubits $n$, gate matrix $U_{2\times2}$ 
    		\Ensure Updated state vector $\psi'$ 
    		\State $mask\_k \gets (1 \ll k)$
    		\For{$i = 0$ \textbf{to} $2^n - 1$}
    		\If{$(i \& mask\_k) = 0$}
    		\State $temp \gets \alpha[i]$
    		\State $\alpha[i] \gets u_{00} \cdot \alpha[i] + u_{01} \cdot \alpha[i + mask\_k]$
    		\State $\alpha[i + mask\_k] \gets u_{10} \cdot temp + u_{11} \cdot \alpha[i + mask\_k]$
    		\EndIf
    		\EndFor
    	\end{algorithmic}
    \end{algorithm}
    
    \begin{algorithm}
    	\caption{CU Gate Operation}\label{algo2}
    	\begin{algorithmic}[1]
    		\Require State vector $\psi$, target qubit $t$, control qubit $c$, total qubits $n$, gate matrix $U_{2\times2}$
    		\Ensure Updated state vector $\psi'$ 
    		\State $mask\_c \gets (1 \ll c)$
    		\State $mask\_t \gets (1 \ll t)$
    		\For{$i = 0$ \textbf{to} $2^n - 1$}
    		\If{$(i \& mask\_c) = 0$}
    		\State \textbf{continue}
    		\EndIf
    		\If{$(i \& mask\_t) = 0$}
    		\State $temp \gets \alpha[i]$
    		\State $\alpha[i] \gets u_{00} \cdot \alpha[i] + u_{01} \cdot \alpha[i + mask\_t]$
    		\State $\alpha[i + mask\_t] \gets u_{10} \cdot temp + u_{11} \cdot \alpha[i + mask\_t]$
    		\EndIf
    		\EndFor
    	\end{algorithmic}
    \end{algorithm}
    
\begin{figure}[ht]
    	\centering
    	\includegraphics[width=\textwidth]{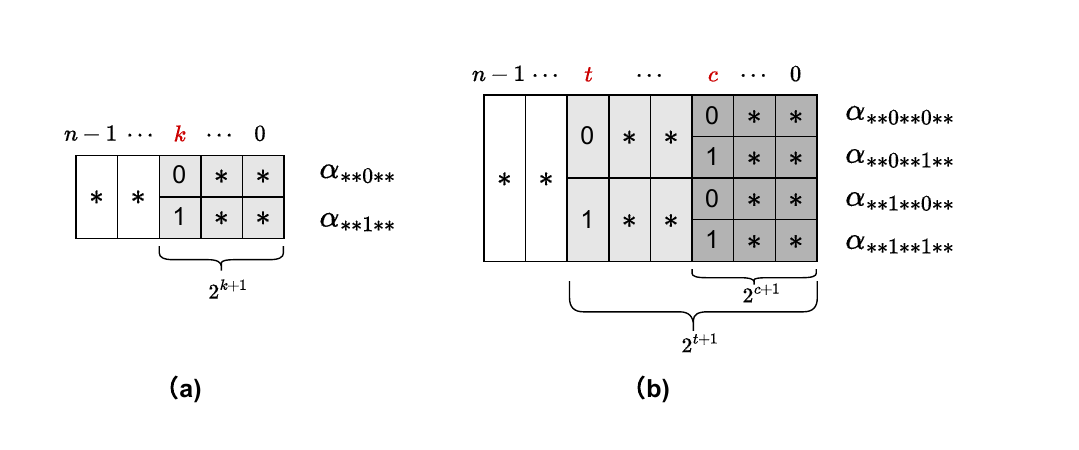}
        \caption{Schematic illustration of the grouping strategy for optimizing single-qubit and control gate operations. (a) For a single-qubit gate acting on qubit $k$, the state vector is divided into groups of size $2^{k+1}$, and only the first half ($2^k$ amplitudes) in each group are traversed, as their paired indices (differing only at the $k$-th bit) can be simultaneously updated. (b) For a control gate with target qubit $t$ and control qubit $c$, the state vector is grouped based on $2^{t+1}$, and within each group, only subgroups where the control bit is 1 (of size $2^c$) are processed. The asterisks ($\ast$) in the quantum state labels denote qubit positions that share the same value in the paired amplitudes.}\label{fig::group_size}
    \end{figure}

    Similar to tiling in matrix-matrix multiplication algorithms, it is also straightforward to further introduce a grouping strategy which can be used to reduce the cost of global traversal and optimize memory locality. As demonstrated in Figure \ref{fig::group_size}, we use the number of coefficients that can be stored in a $k$ ($t$ or $c$) bit vector as a grouping unit. Then, we divide the global quantum state amplitudes into several such units, and traverse half of the amplitudes data in each unit to update the amplitudes in the entire unit. Such an adaptation immediately leads to Algorithms \ref{algo3} and \ref{algo4} given below.
    
    \begin{algorithm}
    	\caption{Single-qubit Gate Operation After Optimized}\label{algo3}
    	\begin{algorithmic}[1]
    		\Require State vector $\psi$, target qubit $k$, total qubits $n$, gate matrix $U_{2\times2}$
    		\Ensure Updated state vector $\psi'$
    		\State $mask\_k \gets (1 \ll k)$
    		\State $group\_size \gets 1 \ll (k+1)$
    		\For{$g = 0$ \textbf{to} $2^n - 1$ \textbf{step} $group\_size$}
    		\For{$i = g$ \textbf{to} $g + mask\_k - 1$}
    		\State $temp \gets \alpha[i]$
    		\State $\alpha[i] \gets u_{00} \cdot \alpha[i] + u_{01} \cdot \alpha[i + mask\_k]$
    		\State $\alpha[i + mask\_k] \gets u_{10} \cdot temp + u_{11} \cdot \alpha[i + mask\_k]$
    		\EndFor
    		\EndFor
    	\end{algorithmic}
    \end{algorithm}
    
    \begin{algorithm}
    	\caption{CU Gate Operation After Optimization}\label{algo4}
    	\begin{algorithmic}[1]
    		\Require State vector $\psi$, target qubit $t$, control qubit $c$, gate matrix $U_{2\times2}$, total qubits $n$
    		\Ensure Updated state vector $\psi'$
    		\State $mask\_c \gets (1 \ll c)$
    		\State $mask\_t \gets (1 \ll t)$
    		\State $group\_size\_c \gets 1 \ll (c + 1)$
    		\State $group\_size\_t \gets 1 \ll (t + 1)$
    		\For{$g = 0$ \textbf{to} $2^n - 1$ \textbf{step} $group\_size\_t$}
    		\For{$s = g + mask\_c$ \textbf{to} $g + mask\_t - 1$ \textbf{step} $group\_size\_c$}
    		\For{$i = s$ \textbf{to} $s + mask\_c - 1$}
    		\State $temp \gets \alpha[i]$
    		\State $\alpha[i] \gets u_{00} \cdot \alpha[i] + u_{01} \cdot \alpha[i + mask\_t]$
    		\State $\alpha[i + mask\_t] \gets u_{10} \cdot temp + u_{11} \cdot \alpha[i + mask\_t]$
    		\EndFor
    		\EndFor
    		\EndFor
    	\end{algorithmic}
    \end{algorithm}

    \section{Parallelism Design for Q\texorpdfstring{$^2$}{2}Chemistry}\label{sec3}

    \subsection{Basic Framework of Distributing a Statevector}\label{subsec3.1}
    
    Following the fundamental algorithms introduced in previous sections, designing a quantum simulator on a single-node or a single GPU device is straightforward, as it only involves nested iterations and shared-memory parallelization, and does not require complex communication designs or synchronization operations between different workers. However, the situation becomes complicated when distributed parallelism is introduced. Specifically, we must address the challenges associated with data communication when distributing the statevector data among multiple processes.
    
    \begin{figure}[ht]
    	\centering
    	\includegraphics[width=0.8\textwidth]{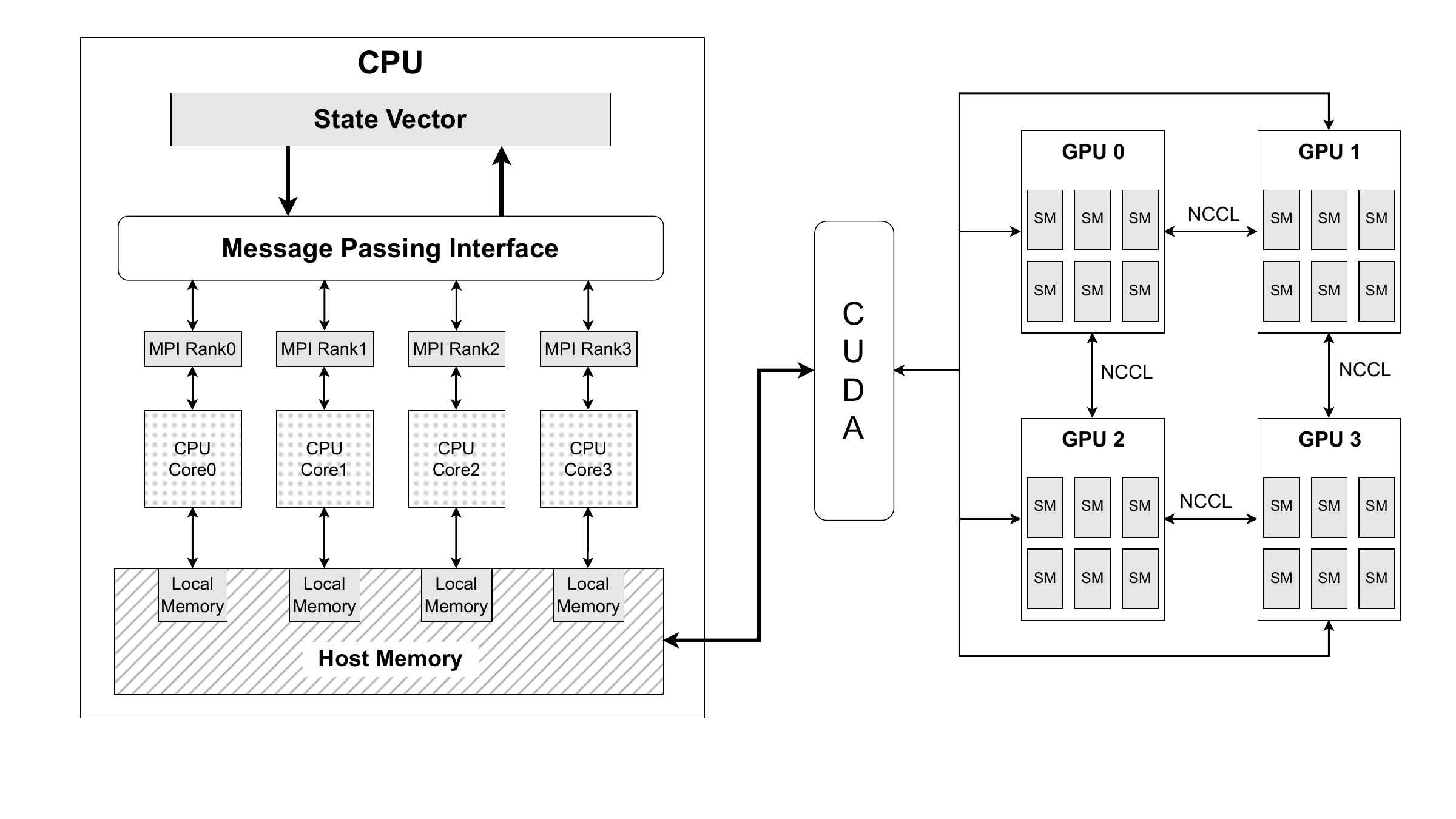}
    	\caption{Quantum state data flow diagram of parallel design of quantum simulator on a parallel system (i.e., heterogeneous multi-GPUs system)}\label{fig::data_flow}
    \end{figure} 
    
    As illustrated in Figure \ref{fig::data_flow}, the quantum state vector is initially constructed in the global memory of the root process and subsequently distributed across multiple workers using MPI (Message Passing Interface). Each MPI process has its own local memory to store a portion of the statevector and tasked to perform computations on its own part. These computations may include applying quantum gates, updating amplitude coefficients and managing data exchanges between ranks. When GPU accelerators are available, data is transferred from host memory (CPU) to device memory (GPU) using CUDA API. Each GPU contains multiple Streaming Multiprocessors (SMs) responsible for handling subsets of amplitudes, ensuring efficient execution of quantum gate operations. Once the computations on the GPUs are completed, the results are transferred back to the CPUs for further processing or final state reconstruction. The final state vector is assembled from all workers.
    
    \begin{figure}[ht]
    	\centering
    	\includegraphics[width=0.8\textwidth]{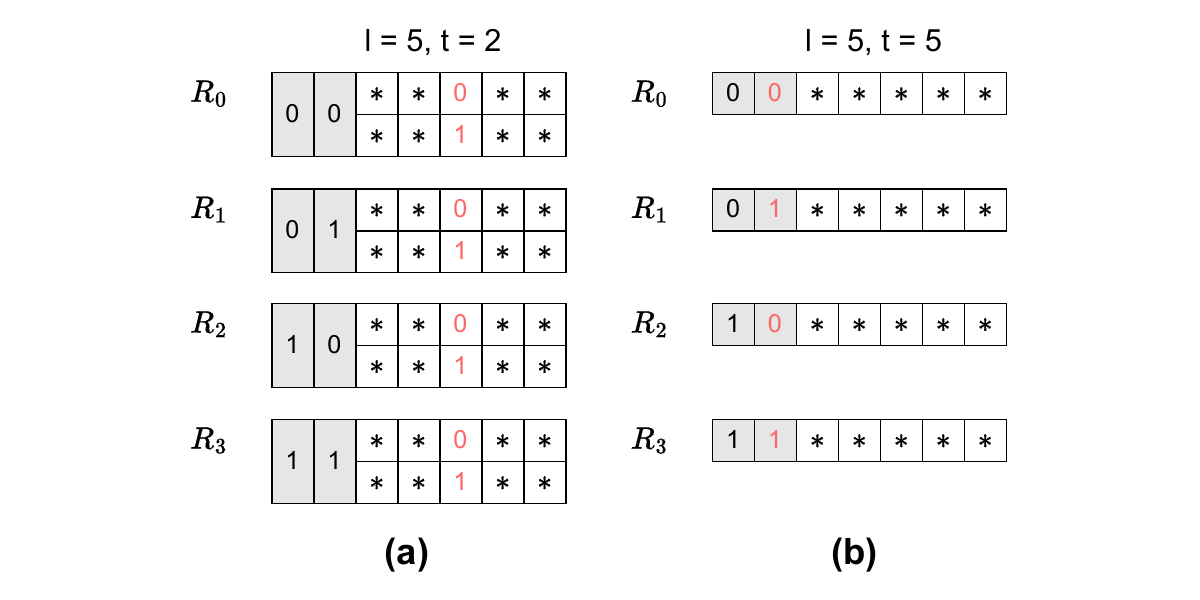}
        \caption{Illustration of how the target qubit position affects the data locality of quantum state amplitudes in a distributed memory setting. (a) when the target qubit index $t$ is less than the number of local qubits $l$, the two amplitudes forming an update pair are reside within the same rank and can be updated locally. (b) In contrast, When the target qubit index $t$ is greater than or equal to the number of local qubits $l$, the two amplitudes are distributed across different MPI ranks, requiring inter-process communication.}\label{fig::rank}
    \end{figure} 
    
    In this distributed parallel design, the need for data communication between processes is determined by the target qubit position and the number of local quantum qubits. Denote $n$ as the total number of quantum qubits. Let the number of processes be $2^m$ and suppose each process holds $2^{n-m}$ coefficients, it can be regarded as each process storing $l=n-m$ qubits. Taking Figure \ref{fig::rank} as an example, a total of $n = 7$ quantum qubits are distributed among $2^{m} = 2^{2} = 4$ processes, with each process stores coefficients for $l = 7-2 = 5$ qubits. It can be observed that the leading $m$ high-qubits of the state correspond directly to the rank of the process, for example, $00$ for rank-$0$ (denoted as $R_{0}$) and $10$ for rank-$2$ ($R_{2}$). Consequently, we can use the high $m$ qubits to represent the process rank which is also recognized as "global qubits", and use the low $l$ qubits to represent "local qubits". The requirement for inter-process communication is governed by the position of the target qubit $t$ with respect to the number of local qubits $l$.
    
    \begin{figure}[ht]
    	\centering
    	\includegraphics[width=0.8\textwidth]{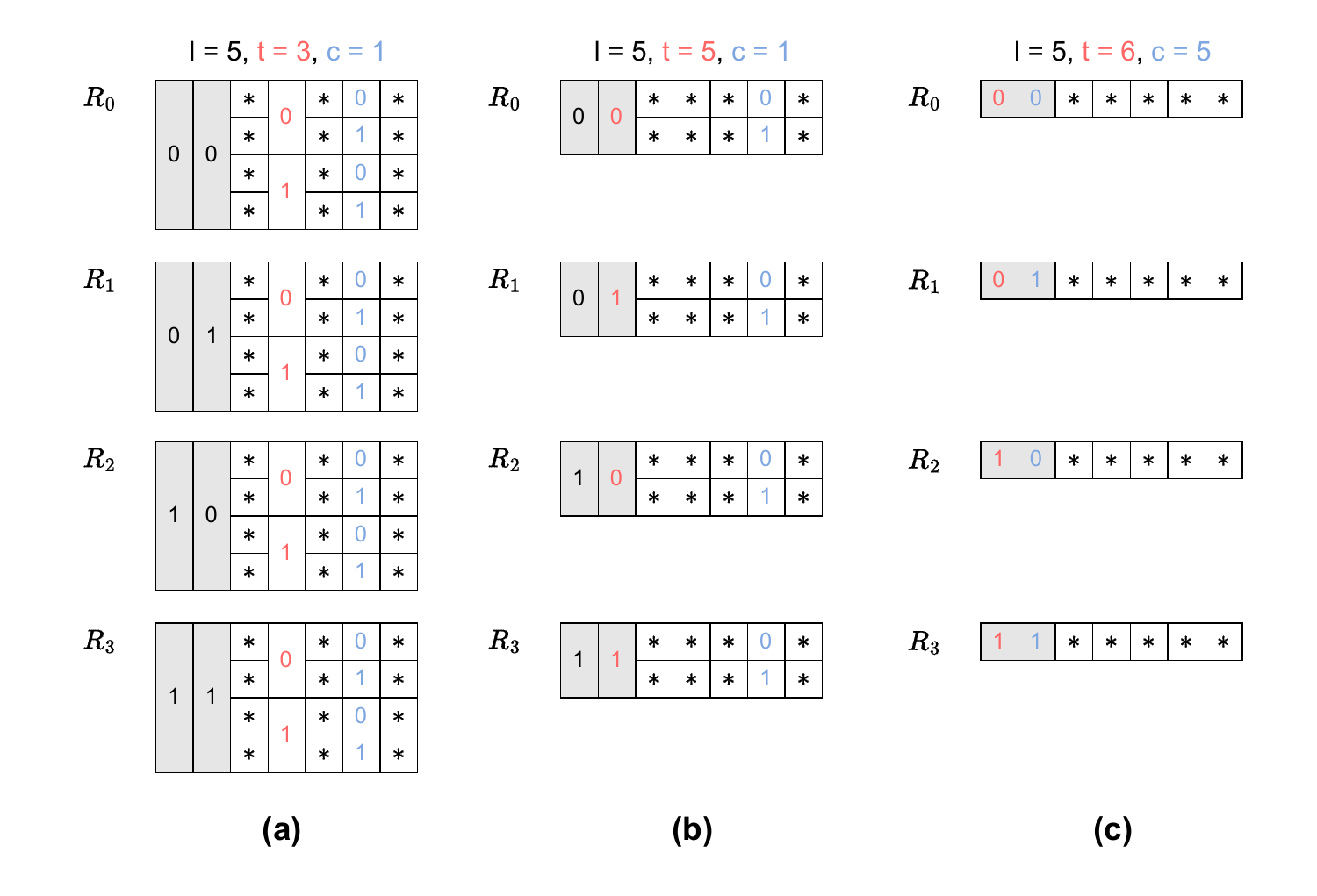}
        \caption{Illustration of the distribution of quantum state amplitude pairs for control gate operations under different configurations of control qubit $c$, target qubit $t$, and the number of local qubits $l$ in a distributed parallel simulation. (a) When $c < t < l$, both the control and target qubits lie within local memory, so no MPI communication is required. (b) When $c < l \le t$ or $t < l \le c$, one qubit remains local, but the other spans multiple ranks, requiring inter-process communication. (c) When $l < c$ and $l < t$, both control and target qubits are distributed, and operations are performed only on processes where the control qubit is set to 1. }\label{fig::rank_cu}
    \end{figure}
    
    For a single-qubit gate, if $l \le t$, let $r$ denote the rank of the process containing the quantum state $|**0_t**\rangle$, the corresponding rank $r_p$ (peer process) containing $|**1_t**\rangle$ can be expressed as:    
    \begin{equation}\label{eq5}
    	r_p = r + \frac{2^t}{2^l} = r \oplus (1 \ll (t-l))
    \end{equation}
    This results in $r = r_p - (1 << (t-l))$. The difference $t-l$ corresponds exactly to the index of the target qubit within the rank of processes. By performing an XOR operation ($\oplus$) between the rank $r$ and $1<< (t-l)$, we can flip the value of the target qubit, changing it from 0 to 1 or from 1 to 0. Therefore, regardless of state (0 or 1) whose amplitude is stored in the current process $r$, we can always find its peer process $r_p$ using the Equation \ref{eq5} and collect the amplitude pair for gate evolution steps.    
    
    For the CU gate, the relationship among the control qubit $c$, target qubit $t$, and the number of local quantum qubits $l$ is illustrated in Figure \ref{fig::rank_cu}. When $c < l \le t$ (Figure \ref{fig::rank_cu}b), this case is identical to the communication scheme used for single-qubit gates with $t \ge l$. In the case of $l \le c < t$ (Figure \ref{fig::rank_cu}c),we only perform operations on those processes where the state of control qubit is 1, the ranks of which can be calculated as:
    \begin{equation*}
    	\centering
    	r \& (1 \ll (c-l))
    \end{equation*}
    
    Within each MPI process, OpenMP-based multi-threading is suggested to accelerate local updates. OpenMP allows multiple CPU threads to concurrently execute gate operations on different segments of the state vector, significantly enhancing throughput. Moreover, we leverage AVX (Advanced Vector Extensions) for vectorized processing of state updates to maximize computation throughput. AVX enables the simultaneous execution of operations on multiple data points at the register level, effectively accelerating matrix-vector multiplications and amplitude transformations.
    In the case of GPU computation where the state vector segments assigned to each MPI process are offloaded to GPU global memory, CUDA kernels are launched to execute gate evolutions in parallel across thousands of GPU threads similar to OpenMP on CPU. GPU-specific libraries can be utilized to optimize inter-GPU communication on multi-GPU platforms, such as NCCL (NVIDIA Collective Communication Library) which efficiently handles data transfer over NVLink, PCIe, or InfiniBand and significantly reduces the communication latency.
    
    Such a distributed parallelism framework for full-amplitude simulation is straightforward and convenient to implement. However, the huge amount of data transfer will introduce a severe performance degradation, and additional memory spaces are required as communication buffer. Suppose an $n$-qubit statevector is distributed among $2^{m}$ processes, where each process is allocated $2^{n-m}$ amplitudes. The amount of data exchanged during gate operations equals to the total number of coefficients handled by a single process, leading to a working memory space of $2^{m-1} \times 2^{n-m}=2^{n-1}$, multiplies the size of one amplitude (usually double-precision complex numbers, i.e., 16 bytes per amplitude). Therefore, the total amount of memory required for simulation will be $3\times 2^{n-1}$ instead of $2^{n}$. In other words, if the system memory can just store $2^{n}$ amplitudes, we can only simulate no more than $n-1$ qubits due to the working space for communication. This is particularly important in the case of GPU computing where GPU memory is generally more limited.
    
    \subsection{Batch-Buffered Overlap Processing}\label{subsec3.2}
\begin{figure}[ht]
    	\centering
    	\includegraphics[width=\textwidth]{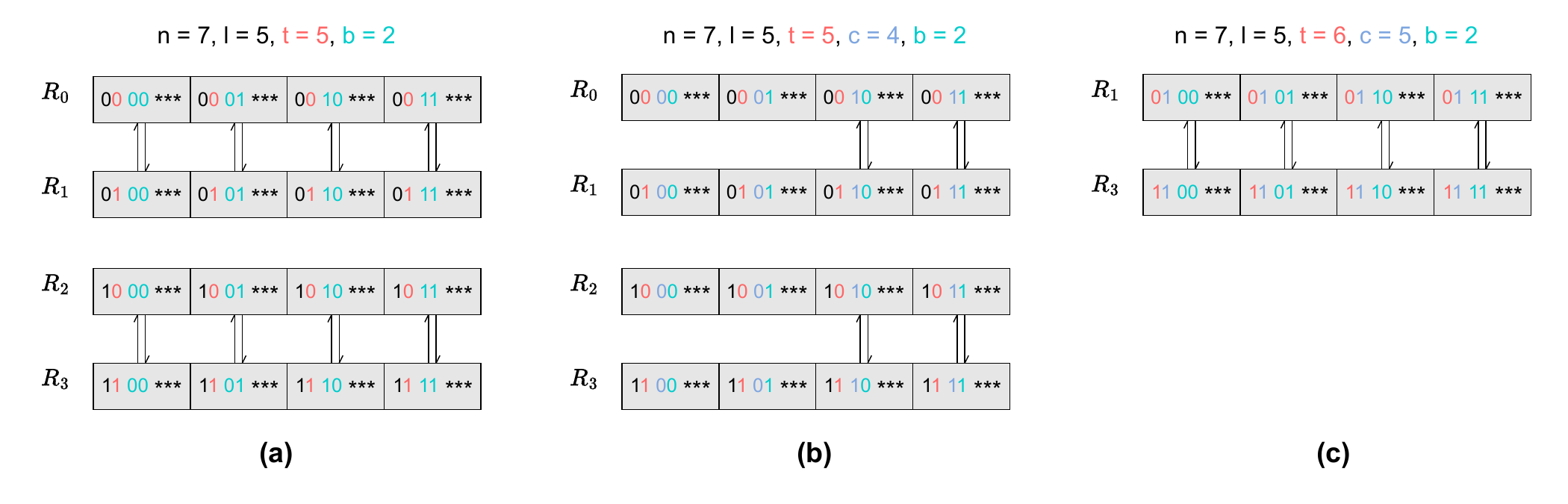}
        \caption{Illustration of inter-process data interaction under batch processing for quantum gate operations in distributed quantum simulation. The batch size is set to $2^b$, allowing each process to operate on a fraction ($1/2^b$) of its local state vector per step. $n$ is the total number of qubits, $l$ is the number of local qubits per process, $t$ is the target qubit index, $c$ is the control qubit index (if applicable), and $b$ denotes the batch granularity. (a) For single-qubit gate with $l \le t$, communication is required across ranks for each batch. (b) For CU gate with $l - b \le c < l \le t$, control qubit resides in local batch space while target qubit spans multiple ranks. (c) When $l \le c < t$, both control and target qubits are distributed, and only the batches where the control qubit equals 1 require communication.}\label{fig::batch}
    \end{figure}

    A useful strategy to lower the overhead of working space is batched processing of gate operations which trades time for space. By setting the batch size to $2^b$ where the high $b$ qubits of local qubits represents the batch index, the system's simulation capability can be increased by $b$ qubits. A batched processing strategy for quantum state evolution calculations within a single process is given in Figure \ref{fig::batch}. The calculation of peer process rank and batch index within the process is similar to the calculation of peer rank in Section \ref{subsec3.1}.
    
    \begin{figure}[ht]
    	\centering
    	\includegraphics[width=0.8\textwidth]{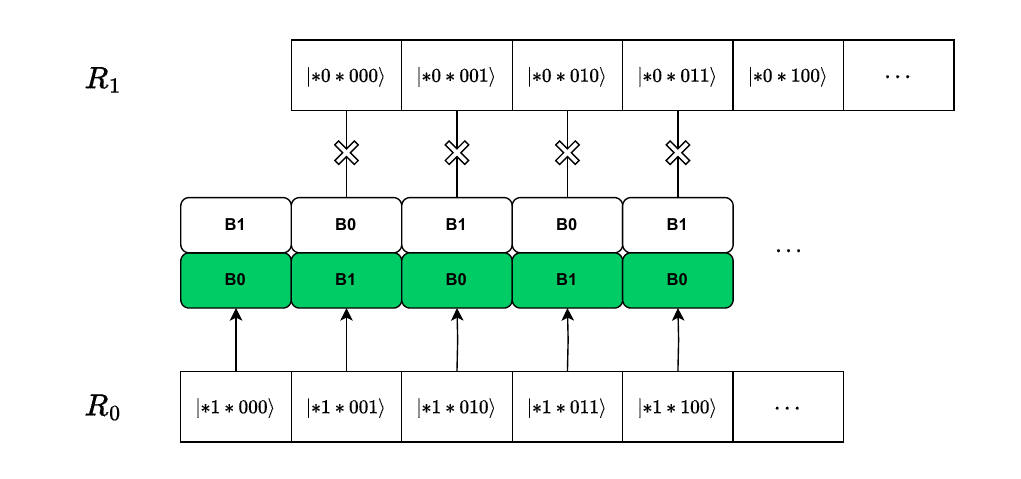}
        \caption{Example of double buffering mechanism for overlapping inter-process communication and computation in distributed quantum simulation. Processes $R_0$ and $R_1$ collaborate on amplitude pair updates, where $R_0$ sends batches such as $|\ast1\ast000\rangle$, $|\ast1\ast001\rangle$, etc., and $R_1$ processes them with corresponding local data like $|\ast0\ast000\rangle$, $|\ast0\ast001\rangle$, using gate matrix operations (denoted by the cross marks). Two memory buffers, B0 and B1 are used alternately in $R_1$ to receive and compute data.}\label{fig::buffer}
    \end{figure}
    
    To further optimize data transfer and computational efficiency in this batched calculation, a multi-buffering strategy is proposed. Assuming that there are only sending, receiving and data calculation operations. A pipeline can be constructed by keeping double buffering in the receiving process. Two memory buffers are maintained: one buffer is actively engaged in computation, the other asynchronously loads the next set of quantum states obtained from the relevant process. As demonstrated in Figure \ref{fig::buffer}, this method significantly covers memory latency and improves overall performance where data movement often becomes a bottleneck.
    
    We name the above strategy as \textbf{Batch-Buffered Overlap Processing (BBOP)}. In a practical implementation, the communication process involves multiple steps, for example: 
    \begin{enumerate}
    \item Process R0 transmitting batch data to the MPI buffer
    \item Process R0 retrieving computed batch data from the MPI buffer to update the original data
    \item Process R1 reading data from the MPI buffer into an local buffer
    \item Process R1 performing computations on the local buffer
    \item Process R1 transferring the updated data back to the MPI buffer
    \end{enumerate}
    Table \ref{tab::buffering} outlines the execution flow of data communication and computation between processes R0 and R1 over a range of time steps (denoted as \textbf{With Buff}), compared with the case that multiple buffers are not implemented (denoted as \textbf{Without Buff}). In the absence of multiple buffers, R0 must wait for R1 to complete the computation of each received batch (denoted as b0, b1, etc.) before transmitting the next batch. This sequential dependency introduces significant idle periods in the workflow. Conversely, the buffered approach effectively hides the communication latency and ensures that the computational process remains uninterrupted. It should be noted that when employing BBOP using non-blocking MPI communication mechanisms, such as \texttt{MPI\_Isend} for sending and \texttt{MPI\_Irecv} for receiving, certain operations can execute implicitly. This allows the pipeline to be automatically asynchronously by the system's data transmission mechanism without too much manual intervention.
    
    \begin{table}[ht]
    	\centering
    	\resizebox{\textwidth}{!}{
    		\renewcommand{\arraystretch}{1.5}
    		\begin{tabular}{c|cccccccc}
    			\hline
    			& T1 & T2 & T3 & T4 & T5 & T6 & T7 & T8\\
    			\hline
    			R0 Without Buff & send b0	&  & &	send b1	& &	 &  send b2&\\
    			\hline
    			R1 Without Buff & &	recv b0	&compute b0	&send b0 back&	recv b1&compute b1&	send b1 back&	recv b2\\
    			\hline
    			R0 With Buff &send b0	&send b1	&send b2	&send b3	&send b4	&send b5	&send b6	&send b7 \\
    			\hline
    			\multirow{3}{*}{R1 With Buff} 	&	&buff0 recv b0	&buff1 recv b1	&buff2 recv b2	&buff0 recv b3	&buff1 recv b4	&buff2 recv b5	&buff0 recv b6\\
    			&	&	&compute b0	&compute b1	&compute b2	&compute b3	&compute b4	&compute b5\\
    			&	&	&	&send b0 back	&send b1 back	&send b2 back&send b3 back	&send b4 back\\
    			\hline
    		\end{tabular}
    	}
    	\caption{Unbuffered and multi-buffered process task scheduling examples}\label{tab::buffering}
    \end{table}
    
    \subsection{Staggered Multi-Gate Parallelism}\label{subsec3.3}
    
    \begin{figure}[ht]
    	\centering
    	\includegraphics[width=0.9\textwidth]{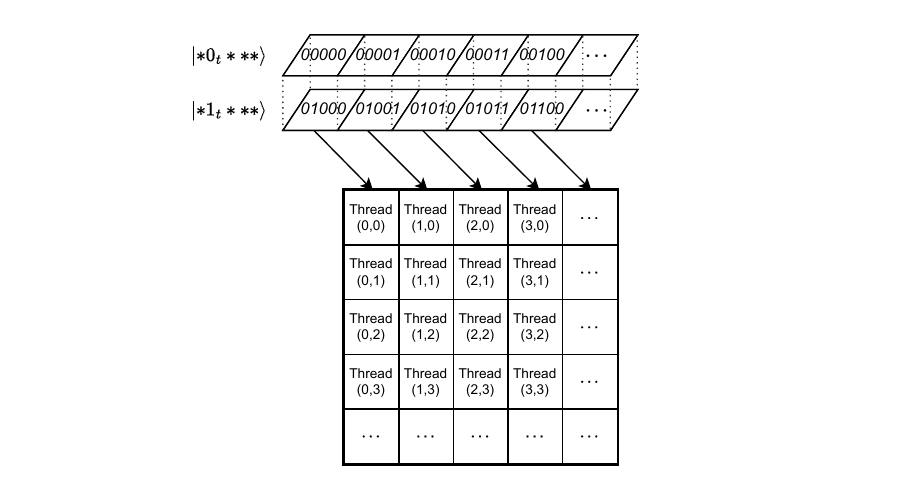}
        \caption{Mapping of quantum state amplitudes to GPU threads in a one-dimensional thread block. Each thread processes a pair of amplitudes corresponding to the target qubit index $t$.}\label{fig::block}
    \end{figure}
    
    GPUs are specifically designed for massively parallel computations, equipped with thousands of cores capable of executing numerous threads. In contrast to CPUs which primarily focus on sequential task execution with complex control logic, GPUs excel in SIMT execution, efficiently managing large-scale data parallelism. When simulating quantum state evolution on a GPU, a one-dimensional block structure is usually employed, where each thread processes a pair of amplitudes independently. As shown in Figure \ref{fig::block}, for a 5-qubit quantum state, when the index of target qubit is 3, pairs of quantum states are mapped one by one to the threads in the block. However, this approach has inherent inefficiencies. Due to the one-dimensional characteristic of quantum states, such data-thread mapping can only utilize a one-dimensional thread block, leading to linear data access which reduces concurrency.
    
    \begin{figure}[ht]
    	\centering
    	\includegraphics[width=0.9\textwidth]{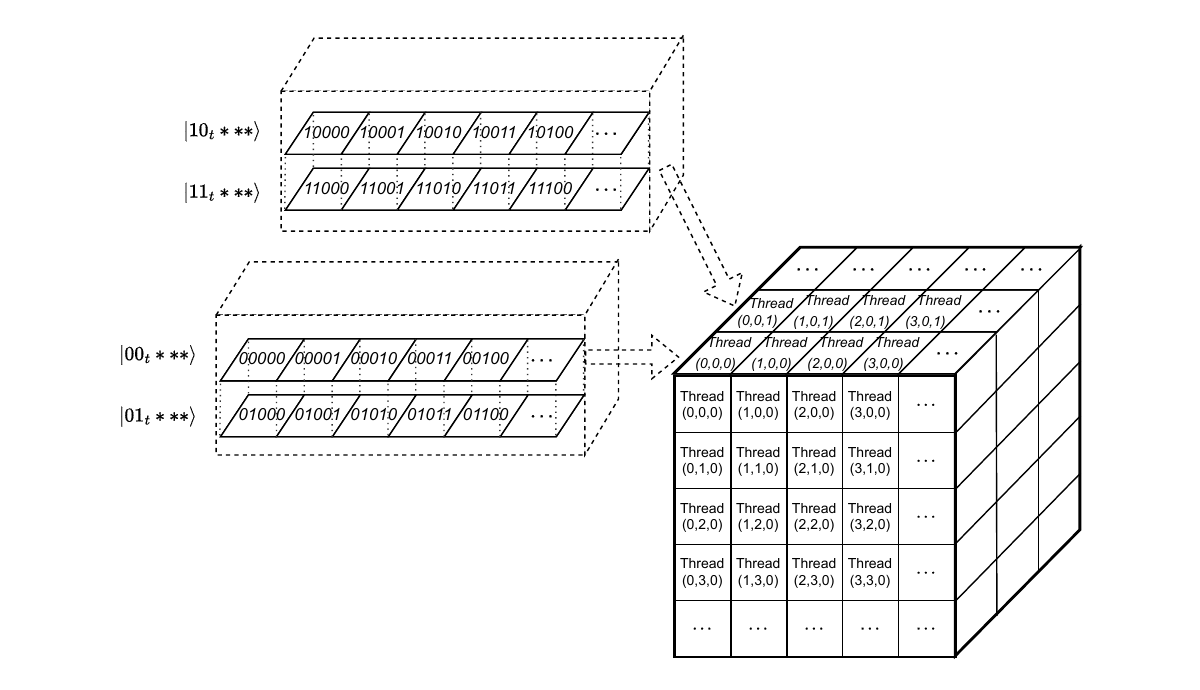}
        \caption{Enhanced two-dimensional block-level thread mapping for staggered execution of quantum gates. By extending the pairing of quantum amplitudes to a two-dimensional grouping scheme.}\label{fig::block_2d}
    \end{figure}
    
    To overcome this limitation, we proposed a two-dimensional block execution model that facilitates the parallel processing of multiple quantum gates, named \textbf{Staggered Multi-Gate Parallelism (SMGP)}. In this approach, the coefficients are mapped onto a 2D thread blocks, making a 3-D topology of overall threading structure. To make use of this two-dimensional block, we group together similar quantum gates that can be executed simultaneously. Figure \ref{fig::block_2d} gives an example of this strategy. The original one-dimensional amplitude pair group $|*0_t\ast\ast\ast\rangle$ and $|*1_t\ast\ast\ast\rangle$ are divided into two-dimensional amplitude pair groups $|00_t\ast\ast\ast\rangle$, $|01_t\ast\ast\ast\rangle$ and $|10_t\ast\ast\ast\rangle$, $|11_t\ast\ast\ast\rangle$, corresponding to statevector updates across different quantum gate calculations. This allows for synchronous calculations on the GPU device while avoiding conflict modification on the same data segment through staggered distribution of quantum state coefficients. 
    
    A detailed timeline of SMGP is given in Table \ref{tab::segment}. Each row corresponds to a different gate $G_i$, while each column $T_i$ represents a time step in the execution process. At each time step, gate $G_i$ applies its transformation to coefficients with a specific set of indices, while the following gates $G_{i+1}$ shifts its target indices cyclically, ensuring that all segments receive updates at different time steps. This approach mitigates conflicts in memory access and enables parallel execution of distinct gate operations on the GPU. Consequently, parallelism scale and computational efficiency are improved via increased utilization of data throughput.
    
    \begin{table}[ht]
    	\centering
    	\caption{Multi-gates perform parallel operations on the same quantum state segment}\label{tab::segment}
    	\begin{tabular}{@{}lllll@{}}  
    		\toprule
    		Time  & T$_0$  & T$_1$  & T$_2$  & T$_3$  \\  
    		\midrule
    		G$_0$  & 00$\ast\ast\ast$  & 01$\ast\ast\ast$  & 10$\ast\ast\ast$  & 11$\ast\ast\ast$  \\  
    		G$_1$  & 01$\ast\ast\ast$  & 10$\ast\ast\ast$  & 11$\ast\ast\ast$  & 00$\ast\ast\ast$  \\  
    		G$_2$  & 10$\ast\ast\ast$  & 11$\ast\ast\ast$  & 00$\ast\ast\ast$  & 01$\ast\ast\ast$  \\  
    		G$_3$  & 11$\ast\ast\ast$  & 00$\ast\ast\ast$  & 01$\ast\ast\ast$  & 10$\ast\ast\ast$  \\  
    		\bottomrule
    	\end{tabular}
    \end{table}
    
    \subsection{Dependency-Aware Gate Contraction}\label{subsec3.4}
    
    Simulation a quantum circuit consists of sequential gate operations. On the top of BBOP and SMGP which optimize individual gate evolution process, the overall computational cost of circuit evolution can be effectively reduced on a higher level. For instance, a single-qubit gate which has a size of $2\times 2$ leads to a total of $2^{n-1}$ matrix-vector multiplications. One such matrix-vector multiplication requires 4 floating-point number multiplications, 2 additions, and 4 data copies. Therefore, assuming uniform time overhead across these operations, the execution time overhead for a single-qubit gate scales as $O(10*2^{n-1})$. Merging two single-qubit gates on different qubits gives a fused two-qubit gate matrix is given in Equation \ref{eq6}. The total operation time overhead of the fusion gate can similarly be calculated as $O(36*2^{n-2})$, which simplifies to $O(9*2^n)$ and becomes smaller than two individual single-qubit gate operations which is $O(2\times 10*2^{n-1}) \rightarrow O(10*2^{n})$.
    
    \begin{equation}\label{eq6}
    	\setlength{\arraycolsep}{2pt}
    	\renewcommand{\arraystretch}{1.5}
    	\begin{array}{rcl}
    		\begin{pmatrix}
    			\alpha'_{*0_{t_2}*0_{t_1}*} \\
    			\alpha'_{*0_{t_2}*0_{t_1}*} \\
    			\alpha'_{*1_{t_2}*0_{t_1}*} \\
    			\alpha'_{*1_{t_2}*1_{t_1}*}
    		\end{pmatrix}
    		& = &
    		U_2 \begin{pmatrix}
    			\alpha_{*0_{t_2}*0_{t_1}*} \\
    			\alpha_{*0_{t_2}*1_{t_1}*} \\
    			\alpha_{*1_{t_2}*0_{t_1}*} \\
    			\alpha_{*1_{t_2}*1_{t_1}*}
    		\end{pmatrix}
    		=
    		\begin{bmatrix}
    			u_{00} & u_{01} & u_{02} & u_{03} \\
    			u_{10} & u_{11} & u_{12} & u_{13} \\
    			u_{20} & u_{21} & u_{22} & u_{23} \\
    			u_{30} & u_{31} & u_{32} & u_{33}
    		\end{bmatrix}
    		\begin{pmatrix}
    			\alpha_{*0_{t_2}*0_{t_1}*} \\
    			\alpha_{*0_{t_2}*1_{t_1}*} \\
    			\alpha_{*1_{t_2}*0_{t_1}*} \\
    			\alpha_{*1_{t_2}*1_{t_1}*}
    		\end{pmatrix}
    		\\
    		& = &
    		\begin{pmatrix}
    			u_{00} \alpha_{*0_{t_2}*0_{t_1}*} + u_{01} \alpha_{*0_{t_2}*1_{t_1}*} + u_{02} \alpha_{*1_{t_2}*0_{t_1}*} + u_{03} \alpha_{*1_{t_2}*1_{t_1}*}\\
    			u_{10} \alpha_{*0_{t_2}*0_{t_1}*} + u_{11} \alpha_{*0_{t_2}*1_{t_1}*} + u_{12} \alpha_{*1_{t_2}*0_{t_1}*} + u_{13} \alpha_{*1_{t_2}*1_{t_1}*}\\
    			u_{20} \alpha_{*0_{t_2}*0_{t_1}*} + u_{21} \alpha_{*0_{t_2}*1_{t_1}*} + u_{22} \alpha_{*1_{t_2}*0_{t_1}*} + u_{23} \alpha_{*1_{t_2}*1_{t_1}*}\\
    			u_{30} \alpha_{*0_{t_2}*0_{t_1}*} + u_{31} \alpha_{*0_{t_2}*1_{t_1}*} + u_{32} \alpha_{*1_{t_2}*0_{t_1}*} + u_{33} \alpha_{*1_{t_2}*1_{t_1}*}\\
    		\end{pmatrix}
    	\end{array}
    \end{equation}
    
    The implementation of quantum gate contraction involves linear algebra calculations and utilizing the properties of quantum gates. For single-qubit gate
    \begin{equation*}
    	M_1 = \begin{bmatrix}
    		\alpha_{00} & \alpha_{01} \\
    		\alpha_{10} & \alpha_{11}
    	\end{bmatrix}
    \end{equation*} 
    acting on target qubit $t_1$ and
    \begin{equation*}
    	M_2 = \begin{bmatrix}
    		\beta_{00} & \beta_{01} \\
    		\beta_{10} & \beta_{11}
    	\end{bmatrix}
    \end{equation*}
    on $t_2$, fusing them into a two-qubit gate $M$ is expressed as the Kronecker product\cite{gate_fusion}:
    \begin{equation}
    	M = M_2 \otimes M_1 =
    	\begin{bmatrix}
    		\beta_{00} & \beta_{01} \\
    		\beta_{10} & \beta_{11}
    	\end{bmatrix}
    	\otimes
            \begin{bmatrix}
    		\alpha_{00} & \alpha_{01} \\
    		\alpha_{10} & \alpha_{11}
    	\end{bmatrix}
    	=
    	\begin{bmatrix}
    		\alpha_{00} \beta_{00} & \alpha_{01} \beta_{00} & \alpha_{00} \beta_{01} & \alpha_{01} \beta_{01} \\
    		\alpha_{10} \beta_{00} & \alpha_{11} \beta_{00} & \alpha_{10} \beta_{01} & \alpha_{11} \beta_{01} \\
    		\alpha_{00} \beta_{10} & \alpha_{01} \beta_{10} & \alpha_{00} \beta_{11} & \alpha_{01} \beta_{11} \\
    		\alpha_{10} \beta_{10} & \alpha_{11} \beta_{10} & \alpha_{10} \beta_{11} & \alpha_{11} \beta_{11}
    	\end{bmatrix}
    \end{equation}
    For adjacent single-qubit gates act on the same qubit (i.e., when $t_1 = t_2$), the result becomes simpler:
    \begin{equation*}
    	M = M_2M_1.
    \end{equation*}
    In this case, the two single-qubit gates effectively merge into a single-qubit gate, maintaining the same $2 \times 2$ matrix size and same target qubit. Similarly, this approach can be extended to CU gates. When both the control qubit and target qubit of two adjacent CU gates CU$_1$ and CU$_2$ are identical, we can also consolidate them into a single CU gate, with $U = U_2 U_1$.
    
    \begin{figure}[ht]
    	\centering
    	\includegraphics[width=1\textwidth]{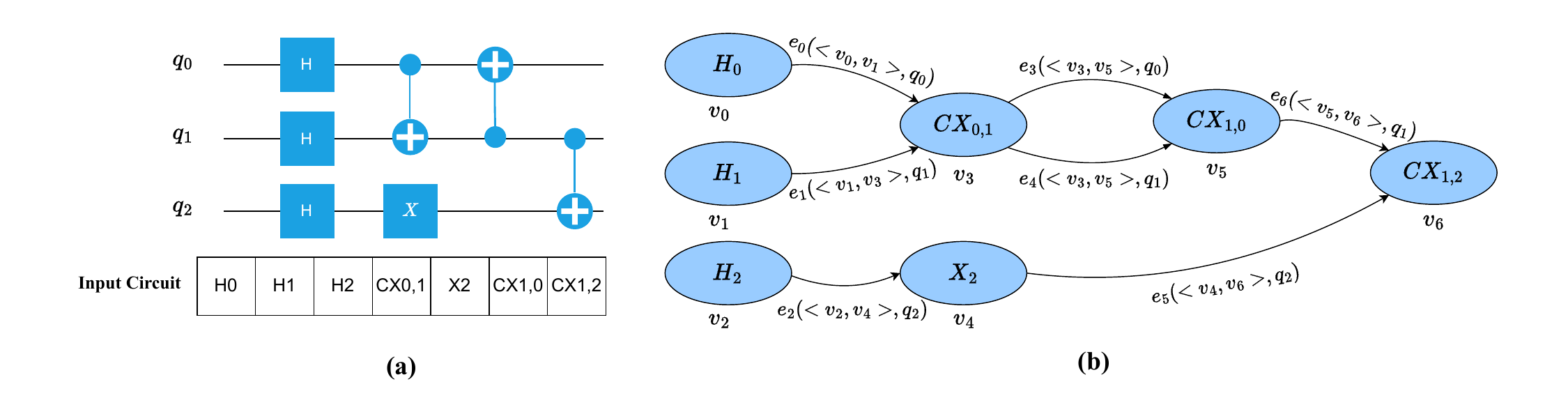}
        \caption{(a) Input quantum circuit where quantum gates are arranged in a layered structure, indicating their application order on qubits $q_0$, $q_1$, $q_2$. (b) Corresponding DAG representation that captures gate-level dependencies. Each vertex represents a quantum gate, and directed edges denote temporal dependencies.}\label{fig::dag_circuit}
    \end{figure}
    
    Since quantum gates in a quantum circuit are time-ordered, to maximize such a contraction process, it is essential to consider the interference relationships among different gate operations. We can represent the quantum circuit using a directed acyclic graph (DAG), where nodes correspond to quantum gates, and directed edges denote dependencies between gates. The DAG construction process involves analyzing qubit operand sets for each gate. For gates $G_i$ and $G_j$ operating on overlapping qubit sets $Q_i \cap Q_j \neq \emptyset$, a dependency edge is established based on program order. Control gates introduce additional dependencies, where a controlled gate $C_{c,t}$ (control qubit $c$, target qubit $t$) depends on all preceding gates affecting either qubit $c$ or $t$. The resulting DAG $\mathcal{G} = (V, E)$ preserves quantum circuit semantics while enabling parallelization analysis. Gates with no path between them in the DAG can execute concurrently, facilitating optimal resource utilization and circuit depth (in terms of layers instead of gate counts) reduction. Figure \ref{fig::dag_circuit}
    shows a quantum circuit that uses a DAG structure to encode gate dependencies and execution order. After the DAG is constructed, A greedy algorithm is employed to merge quantum gates. This DAG-based gate contraction method is named as \textbf{Dependency-Aware Gate Contraction (DAGC)}.

    \section{Results}\label{sec4}

    \subsection{Experiment Setup}\label{subsec4.1}
    
    The optimized Q$^{2}$Chemistry full-amplitude quantum simulator is benchmarked in both CPU-only cluster and GPU heterogeneous environments. We conduct various performance evaluations to validate the generality and scalability of our optimization scheme.
    
    The CPU-only cluster is a multi-node system with each node equipped with two 64-core AMD EPYC 7763 CPUs, providing a total number of 128 threads per node. It has a base frequency of 2.45 GHz and supports 8-channel DDR4 memory with 204.8 GB/s bandwidth. InfiniBand is used for inter-node connection, offering a bandwidth of 100 Gbps $\times$ 4 lanes per port. The GPU heterogeneous system is a single node equipped with two 24-core AMD EPYC 7402 CPUs and 8 NVIDIA A100 GPUs connected using PCI-E channels. Each A100 GPU includes 6,912 CUDA cores, 432 Tensor Cores, and 40 GB of HBM2 memory offering a 1,555 GB/s intra-card memory I/O performance.
    
    On the software side, we utilize Intel's MPI software package, version 2021.3.0, which supports compilation and execution in an MPI+OpenMP environment, enabling efficient hybrid parallelism across distributed and shared memory architectures. The GPU kernels are built using NVIDIA's CUDA version 12.1. We compile the program using mpicxx and nvcc that come with MPI and CUDA, maintaining consistent versions.
    
    Three types of circuits are used in our experiments, including a Quantum Fourier Transform (QFT) circuit, a Quantum Approximate Optimization Algorithm (QAOA) circuit and a hardware-efficient ansatz (HEA) circuit which includes five layers of parameterized single-qubit gates (RX, RY, RZ) and staggered CNOT gates. All circuits are scaled to 20 to 30 qubits, with the total number of circuit gates ranges from 238 to 1652. We employ the qHiPSTER circuit format as the reference design for the circuit input file, which retains all quantum gate elements in the OpenQASM 2.0 format.
    
    \subsection{Scalability Analysis}\label{subsec4.2}
    \begin{figure}[ht]
    	\centering
    	\includegraphics[width=0.8\textwidth]{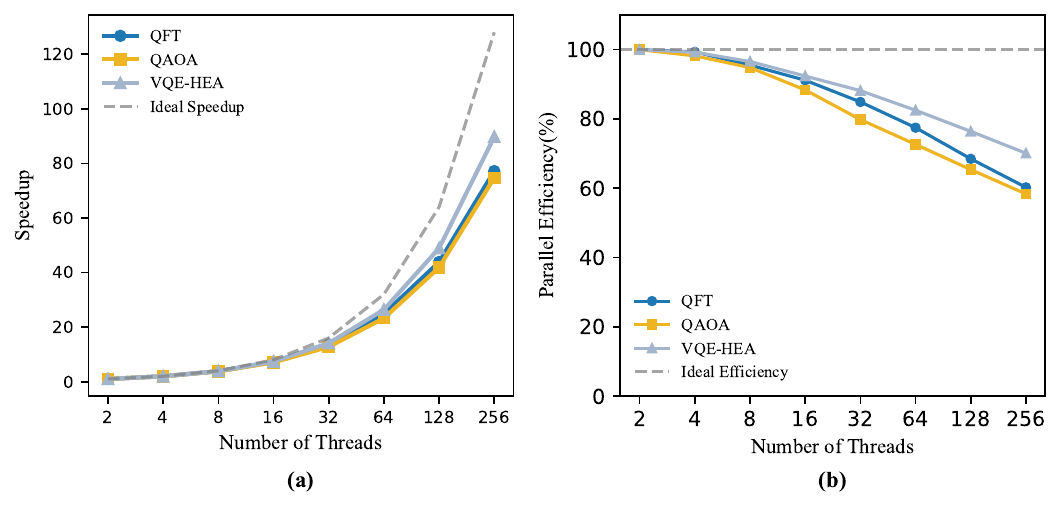}
        \caption{Strong scaling performance of 28-qubit quantum circuit simulation on CPU cluster. (a) Speedup relative to 2-thread baseline shows QFT, QAOA, and VQE-HEA achieving 77.1×, 74.7×, and 89.8× speedup respectively at 256 threads. (b) Parallel efficiency demonstrates excellent scaling up to 32 threads ($>$80\% efficiency), with degradation beyond 64 threads due to multi-node communication overhead and thread allocation inefficiencies.}\label{fig::strong_scaling_cpu}
    \end{figure}
    
    Strong scaling were tested using the 28-qubit quantum circuits for CPU and GPU. In the CPU-only case, we use OpenMP for tests below 64 threads and MPI plus OpenMP for above, with the number of OpenMP threads per process set to 64. Weak scaling tests were conducted using QAOA circuits. On CPU clusters, we use four different thread configurations in a staggered manner: 4 threads for 22-28 qubits, 8 threads for 23-29 qubits, 16 threads for 24-30 qubits, and 32 threads for 25-31 qubits. Similarly, three configurations (1, 2 and 4 GPUs) are implemented in the case of GPU benchmark.

\begin{figure}[ht]
        \centering
        \includegraphics[width=0.8\textwidth]{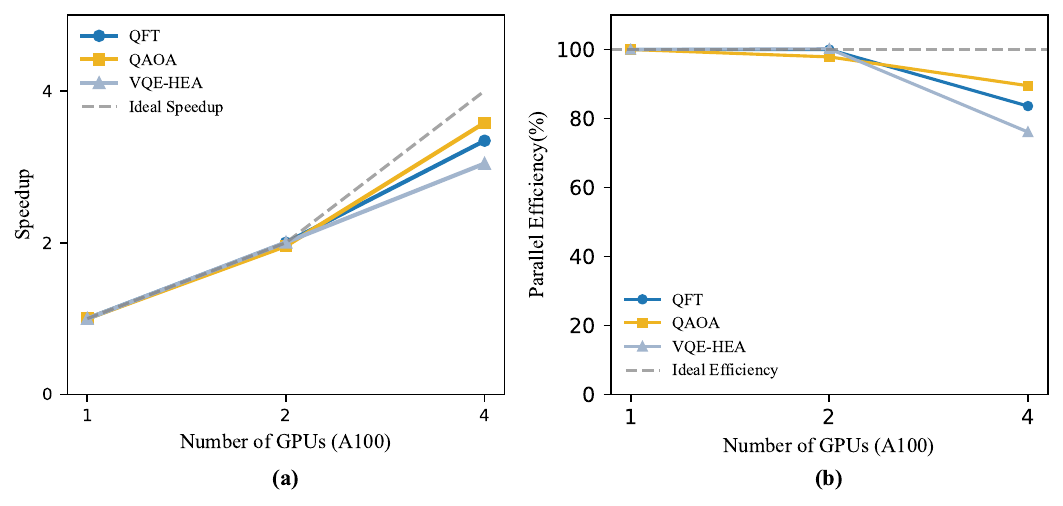}
        \caption{Strong scaling performance of 32-qubit quantum circuit simulation on multi-GPU cluster. (a) Speedup relative to single GPU increases significantly with additional GPUs for QFT, QAOA, and VQE-HEA circuits, reaching $3.34\times$, $3.58\times$, and $3.04\times$ with 4 GPUs. (b) When scaling to 4 GPUs, circuits with larger kernel workloads (QFT, QAOA) maintain higher efficiency (83.5\%, 89.5\%), while VQE-HEA, whose kernel computation is relatively lightweight, exhibits the most significant efficiency drop, reaching about 76\%.}
        \label{fig::strong_scaling_gpu}
    \end{figure}

    Figure \ref{fig::strong_scaling_cpu} and Figure ~\ref{fig::strong_scaling_gpu} demonstrates the parallel efficiency for CPU and GPU respectively. Near ideal scaling is obtained for small-scaled parallelism (e.g., 8 threads or 2 GPUs), while the efficiency drops with increased computation scale. This is mainly caused by load imbalancing, where quantum gate operations targeting specific qubit pairs tend to create irregular computational patterns, leading to more unevenly distributed worker tasks with increased number of workers.
    
\begin{figure}[ht]
    	\centering
    	\includegraphics[width=0.8\textwidth]{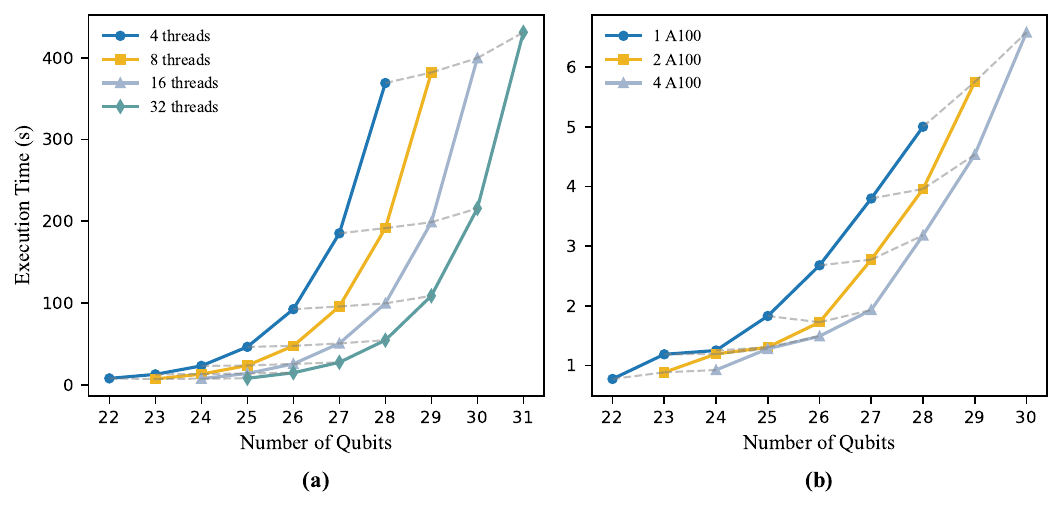}
    	\caption{(a) Performance of QAOA circuit simulation on a CPU cluster. In such thread configurations (22 qubits @ 4 threads to 31 qubits @ 32 threads), the computational overhead on each thread is kept the same. Dashed lines connect configurations with equivalent per-thread workload. (b) Performance of QFT circuits simulation on multi-GPU system. Execution time is measured with 1, 2, and 4 A100 GPUs based on NCCL communication.}\label{fig::weak_scaling}
    \end{figure}

    The CPU-only weak scaling benchmark is given in Figure \ref{fig::weak_scaling}a. Generally, all four thread configurations exhibit similar execution time as the qubit number grows, indicating that the parallel simulator sustains stable performance as both the problem size and thread count increase. A closer comparison reveals that the average work efficiency per thread slightly decreases as the thread count increases (i.e., execution time slightly increases), as indicated by the dashed lines connecting horizontally the data points which are expected to have identical computational overhead. As the number of threads increases, the uneven mapping of amplitude pairs to threads involved in gate updates leads to more threads being idling and consequently degrades parallel performance. This is also observed in the case of multi-GPU (Figure \ref{fig::weak_scaling}b), where the number of threads launched by computational kernels on a GPU is orders of magnitude larger than on a CPU.
    
    \subsection{Ablation Studies}\label{subsec4.3}
    \begin{figure}[ht]
    	\centering
    	\includegraphics[width=0.8\textwidth]{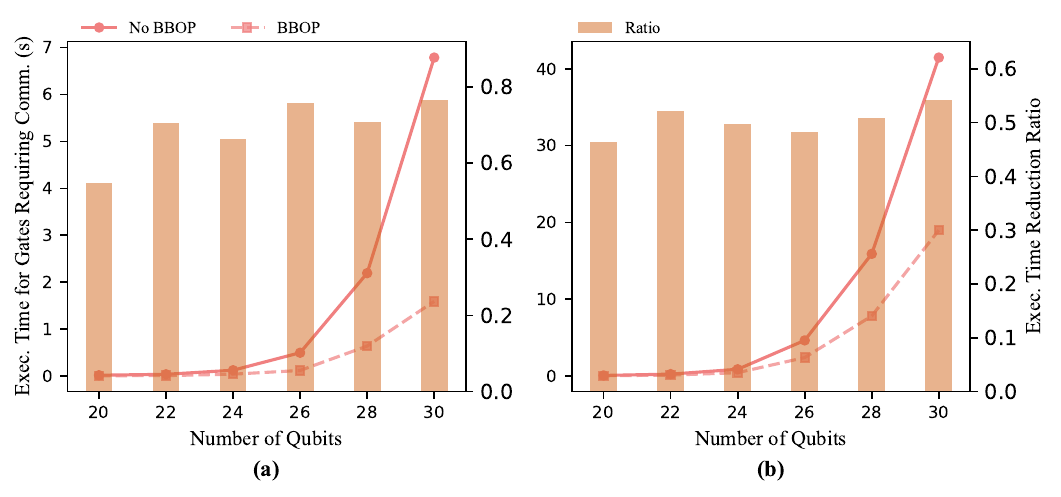}
        \caption{Execution time comparison for communication-requiring quantum gates before and after applying the BBOP strategy. (a)VQE-HEA circuits and (b) QAOA circuits tests with various qubit number, using 4 MPI processes and 16 OpenMP threads per process on a CPU-only cluster. BBOP reduces inter-process communication overhead, especially in VQE-HEA circuits, achieving up to 76.5\% time reduction at 30 qubits.}\label{fig::BBOP_cpu}
    \end{figure}
    
    \textbf{Batch-Buffered Overlap Processing} We first conducted BBOP (Batch-Buffered Overlap Processing) strategy across different quantum circuit configurations with CPU-only cluster. As shown in Figure \ref{fig::BBOP_cpu}, for the VQE-HEA configuration, the total execution time for quantum gates that requires inter-worker communication is significantly reduced with BBOP optimization (69.08\% on average across all tested qubit sizes ). Similarly, for the QAOA configuration, BBOP optimization achieves an average reduction rate of 50.23\%, highlighting its efficiency in minimizing communication overhead.
    
    In contrast, GPU implementations revealed BBOP's limited effectiveness due to hardware-specific machine balance (peak FLOPs divided by peak bandwidth). Table \ref{tab:bbop_gpu} shows that inter-process communication dominates execution time (approximately 99\%) for high-order qubit operations, while actual computation accounts for merely 0.5\%-1\%. This prevents effective overlap between communication and computation in GPU environments, and the 2.66× speedup observed in CPU clusters diminishes to negligible levels when transitioning to GPU acceleration. The fundamental difference arises from GPU platform's performance characteristic. While they execute gate operations 1-2 orders of magnitude faster than CPUs, their reliance on collective communication patterns through NCCL via PCI-E links creates data transfer bottlenecks that cannot be easily mitigated by multi-buffering alone. These findings suggest BBOP should be selectively applied based on hardware specifications. It is crucial for CPU clusters with distributed memory architectures, but becomes less effective for GPUs' such communication-bound nature in heterogeneous systems. 
    

    \begin{table}[h]
    	\centering
    	\caption{BBOP Performance Analysis on 4 GPU Devices for VQE-HEA Circuit. $T_{commun}$ and $T_{comput}$ refers to time cost for communication and computation respectively.}
    	\label{tab:bbop_gpu}
    	\begin{tabular}{c|ccc}
    		\hline
            \textbf{Qubits} & \textbf{$T_{commun}$ (s)} & \textbf{$T_{comput}$ (s)} & \textbf{Communication Ratio} \\
            \hline
            20 & 0.06754 & 0.00065 & 99.03\% \\
            22 & 0.84269 & 0.00087 & 99.90\% \\
            24 & 0.95176 & 0.00166 & 99.83\% \\
            26 & 1.38303 & 0.00404 & 99.71\% \\
            28 & 2.91954 & 0.01604 & 99.45\% \\
            30 & 6.83090 & 0.07450 & 98.91\% \\
            \hline
            
    		\hline
    	\end{tabular}
    \end{table}
    
    \begin{figure}[ht]
    	\centering
    	\includegraphics[width=0.8\textwidth]{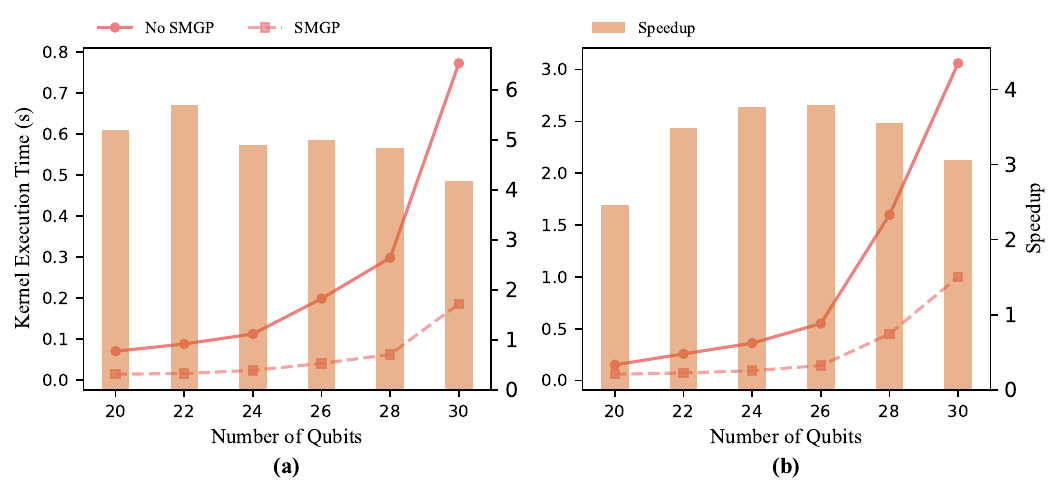}
         \caption{Execution time speedup achieved through SMGP (Staggered Multi-Gate Parallelism) for (a) QAOA and (b) VQE-HEA circuits on GPU-based simulations. Average speedups are 3.35× (QAOA) and 4.96× (VQE-HEA).}\label{fig::SMGP_speedup} 
    \end{figure}

    \textbf{Staggered Multi-Gate Parallelism} 
    Building on our previous analysis of communication bottlenecks in GPU environments, this section presents SMGP as a complementary optimization for BBOP that specifically addresses the concurrency limitations in GPU calculations. The strategy redefines thread block organization through two-dimensional amplitude grouping, transforming memory access patterns to maximize GPU throughput while minimizing inter-block synchronization. 
    
    \begin{figure}[ht]
    	\centering
    	\includegraphics[width=0.8\textwidth]{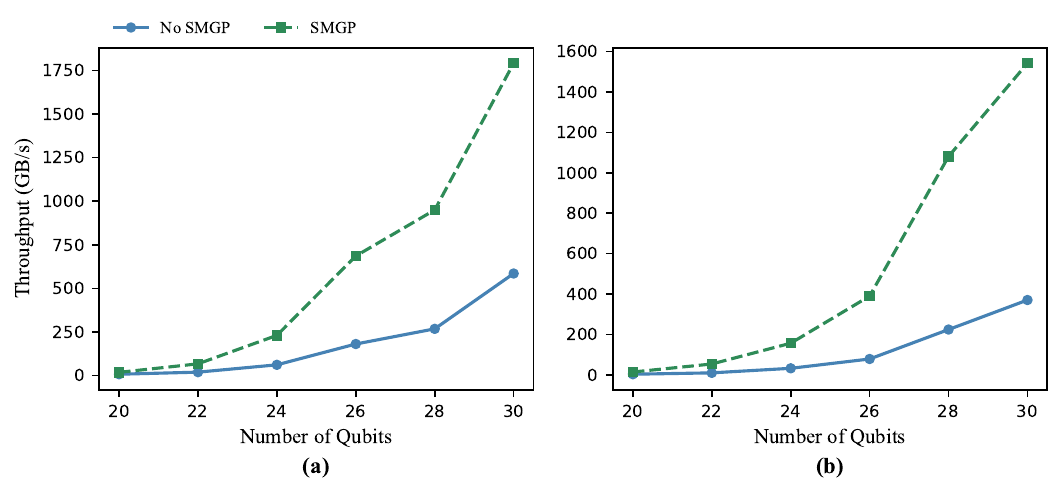}
        \caption{Improvement in memory throughput before and after applying SMGP for QAOA and VQE-HEA circuits. Average throughput increases from 186.86 GB/s to 623.71 GB/s for (a) QAOA circuit, and from 119.25 GB/s to 539.13 GB/s for (b) VQE-HEA circuit, with peak values reaching up to 1791.57 GB/s and 1541.95 GB/s.}\label{fig::SMGP_throughput} 
    \end{figure}
    
    As shown in Figure \ref{fig::SMGP_speedup}, SMGP achieves substantial execution time reductions across both circuit types when implemented on A100 GPUs. While QAOA circuits maintain regular gate patterns amenable to moderate parallelism, VQE-HEA's alternating layers of RX/RY/RZ gates and CNOT operations create more opportunities for concurrent execution within the staggered block structure.
    
    The throughput analysis in Figure \ref{fig::SMGP_throughput} reveals SMGP's memory optimization capabilities during quantum state evolution by optimizing memory access patterns. As shown in Figure~\ref{fig::SMGP_throughput}, SMGP yields 3.3× and 4.5× improvements in average memory throughput for QAOA and VQE-HEA circuits, respectively. These results confirm that SMGP effectively increases memory-level parallelism by maximizing concurrent access and reducing idle time during large-scale sequential gate operations.
    
    \begin{figure}[ht]
    	\centering
    	\includegraphics[width=0.8\textwidth]{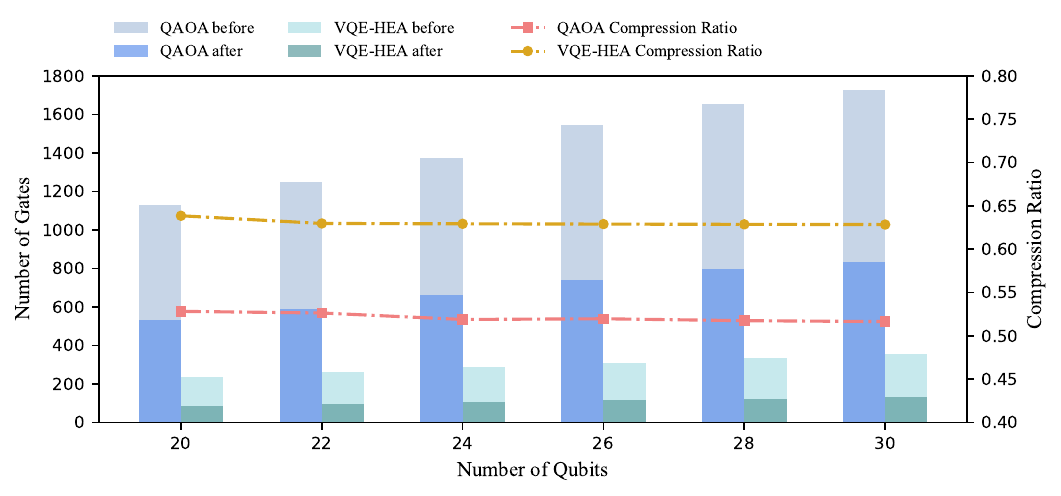}
        \caption{Reduction in the number of communication-related quantum gates before and after applying DAGC. Results are shown for QAOA and VQE-HEA circuits. The compression ratio is computed as the percentage of reduced gates relative to the original number, i.e., $(N_{\text{before}} - N_{\text{after}})/N_{\text{before}} \times 100\%$. The average compression rates are 52.13\% for QAOA and 63.07\% for VQE-HEA circuits.}\label{fig::DAGC_compression}
     \end{figure}
    
    \textbf{Dependency-Aware Gate Contraction} 
    While previous optimizations focused on enhancing memory bandwidth utilization and parallel execution efficiency at the individual gate level, DAGC works as a structural optimization that fundamentally reduces gate operation counts through circuit-level dependency analysis. Figure \ref{fig::DAGC_compression} illustrates the significant gate count reduction achieved by the structural compression across QAOA and VQE-HEA benchmarks. Such an effectiveness has strong correlation with circuit topology characteristics. VQE-HEA circuits achieve superior compression rates compared to QAOA due to their higher proportion of parameterized single-qubit gates. 
    
\begin{figure}[ht]
    	\centering
    	\includegraphics[width=0.8\textwidth]{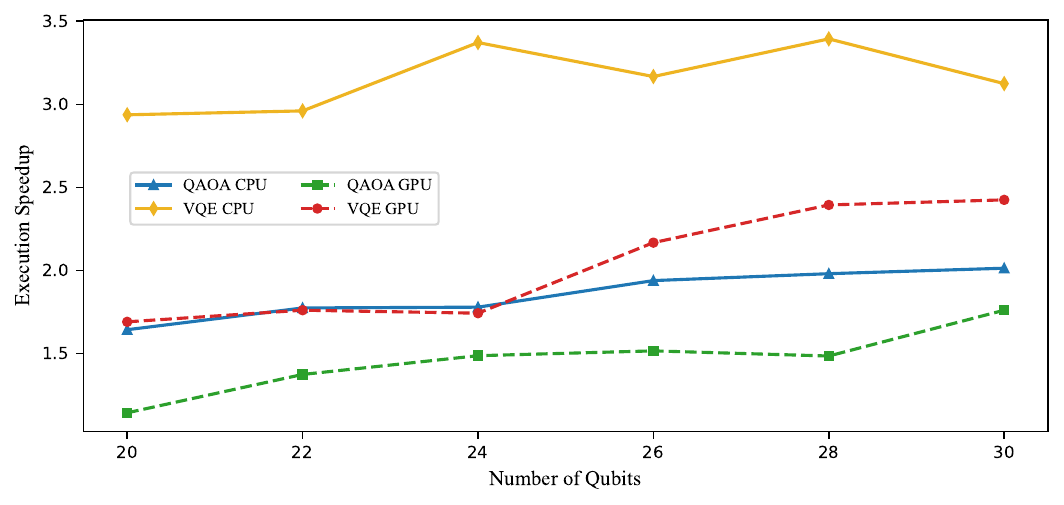}
        \caption{Execution time speedup achieved by applying DAGC to QAOA and VQE-HEA circuits across two hardware settings: a CPU-only cluster and the heterogeneous GPU platform. Speedup is computed as the ratio of execution time before and after applying DAGC.}\label{fig::DAGC_speedup}
    \end{figure}

    As presented in Figure \ref{fig::DAGC_speedup}, the resulting gate count reduction directly leads to execution time improvements across both hardware platforms, with VQE-HEA simulations showing 3.15× speedup in CPU-only environments and 2.03× acceleration in heterogeneous systems. QAOA benchmarks demonstrate comparable but slightly reduced gains (1.85× and 1.46× for CPU and heterogeneous systems respectively), reflecting the algorithm's more constrained gate sequence that limits fusion potential. These results validate DAGC's dual benefits that it not only reduces computational complexity through matrix size optimization, but also indirectly minimizes communication requirements by decreasing the number of distributed gate operations. 
    
\begin{figure}[ht]
    	\centering
    	\includegraphics[width=0.8\textwidth]{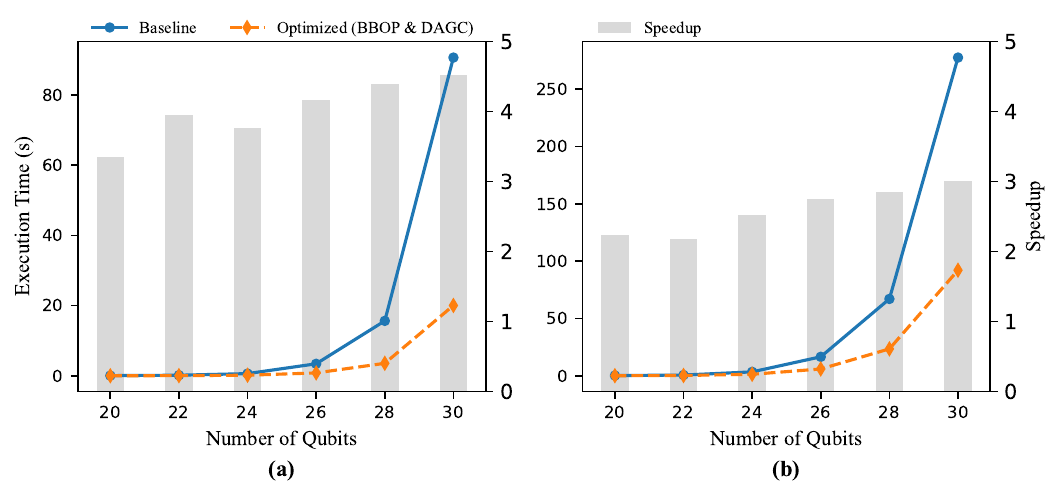}
    	\caption{Execution time and corresponding speedup for (a) VQE-HEA and (b) QAOA circuits on CPU-only cluster before and after applying combined optimizations (BBOP + DAGC).}\label{fig::fullopt_cpu}
    \end{figure}

\begin{figure}[ht]
    	\centering
    	\includegraphics[width=0.8\textwidth]{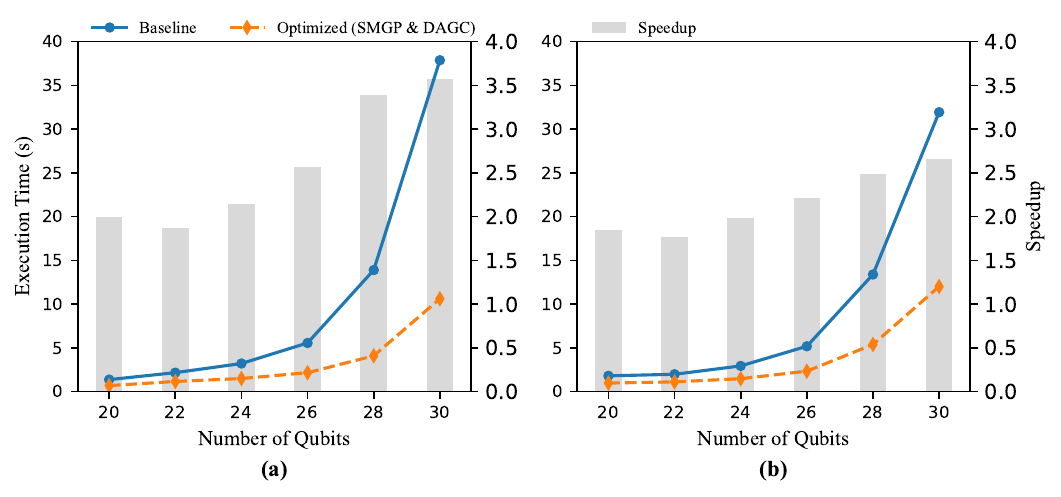}
    	\caption{Execution time and corresponding speedup for (a) VQE-HEA and (b) QAOA circuits on heterogeneous GPU platform before and after applying combined optimizations (SMGP + DAGC).}\label{fig::fullopt_gpu} 
    \end{figure}

    Finally, we conducted comprehensive performance evaluation by applying all valid optimization strategies (BBOP plus DAGC for CPU-only environment and SMGP plus DAGC for GPU system) across both hardware architectures. Figure \ref{fig::fullopt_cpu} and \ref{fig::fullopt_gpu} demonstrate the synergistic effects of these optimizations through 20-30 qubit benchmarks with 64 threads and four A100 GPUs respectively. For CPU-only execution, VQE-HEA circuits achieve an overall of 4.52× speedup (90.652s $\rightarrow$ 20.038s at 30 qubits) while QAOA shows 3.01× improvement (277.357s $\rightarrow$ 92.084s). The GPU implementation show a similar performance gain, delivering 3.57× and 2.66× speedups for VQE-HEA and QAOA circuits at 30 qubits respectively. Notably, the optimization effectiveness scales with qubit count in both environments, with 30-qubit simulations showing greater acceleration than 20-qubit benchmarks. 
    
    The observed performance trends reveal critical architecture-specific limitations and optimization opportunities. CPU clusters benefit most from BBOP's communication-computation overlap (76.52\% execution time reduction in communication-bound gates, Figure \ref{fig::BBOP_cpu}) and DAGC's structural compression (63.07\% gate count reduction in VQE-HEA, Figure \ref{fig::DAGC_compression}). However, the scalability of these optimizations is constrained by Amdahl's law in distributed-memory architectures, where synchronization overhead grows with core count. In contrast, GPU acceleration demonstrates superior scalability for compute-bound operations through SMGP's two-dimensional thread blocks, achieving 4.96× speedup in VQE-HEA simulations. The combination of gate fusion and staggered execution patterns enables 1.84×-3.57× total speedups while maintaining memory efficiency through optimized register allocation. These findings suggest that architecture-specific optimizations should be implemented: GPU favors SMGP's efficient thread modeling for concurrency complemented with DAGC gate compression to achieve optimal performance, while CPU clusters require BBOP's pipelined execution to mitigate communication overhead in large-scale simulations. 
    
    \subsection{Cross-software Benchmark}\label{subsec4.4}
    To further assess the practical advantages of our quantum simulator, we compare it against leading open-source simulators (QuEST, Qulacs, ProjectQ, Qsim, Qiskit, MindQuantum, Pennylane, Qibo and Yao) under both CPU-only and GPU heterogeneous architectures. The benchmark follows the same hardware configurations detailed in Section~\ref{sec4}: 4-node CPU clusters with 64 cores (Qiskit, Yao, MindQuantum, and Qibo use 64 threads in a single node, while others use 4 MPI × 16 OpenMP threads) and GPU systems with 1-4 A100 devices. All tests employ standardized circuits (QFT, VQE-HEA, QAOA) with qubit counts ranging from 20 to 30, ensuring consistent evaluation across different software stacks. 
    
    \begin{figure}[ht]
        \centering
        \includegraphics[width=0.8\textwidth]{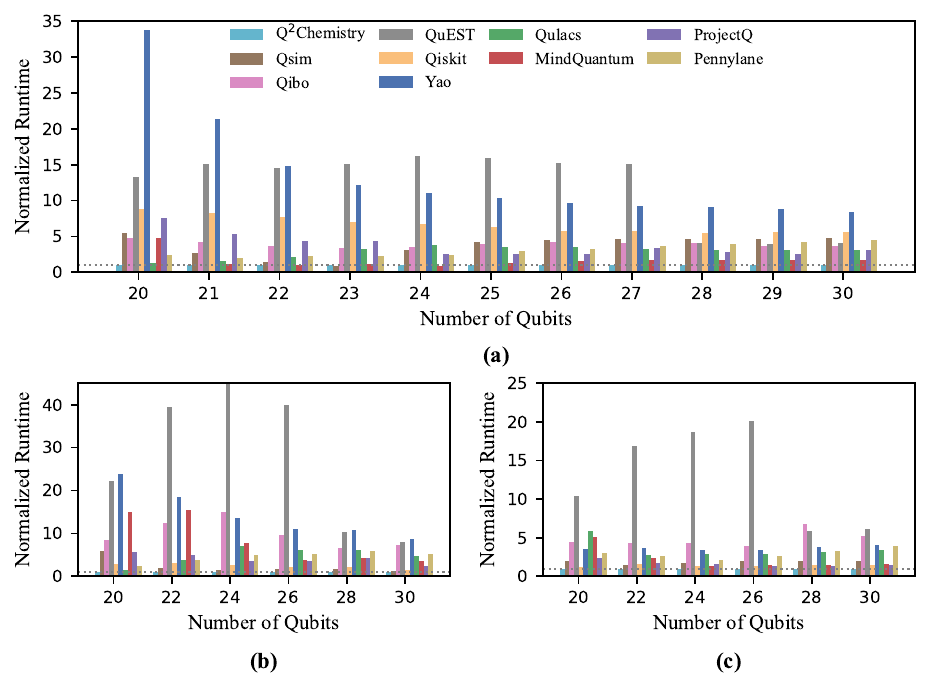}
        \caption{Normalized execution time comparison of (a) QFT, (b) VQE-HEA, and (c) QAOA circuit simulations on a 64-thread CPU cluster. Results are normalized to Q$^2$Chemistry performance (baseline = 1, shown as dashed horizontal line).}
        \label{fig:cpu-benchmark}
    \end{figure}
    
    As shown in Figure~\ref{fig:cpu-benchmark}, Q$^2$Chemistry demonstrates dominant performance in CPU-only environments. For QFT circuits, Q$^2$Chemistry outperforms established simulators such as Qiskit and Pennylane, reaching up to 13.50× faster than Yao and 12.06× faster than QuEST. In the VQE-HEA benchmark, similar superiority is observed with a maximum performance gain up to 45.01x. In QAOA circuits, where performance variations tend to be smaller, Q$^2$Chemistry still demonstrates solid gains. On average, Q$^2$Chemistry achieves overall speedups of 7.91× for QFT, 11.46× for VQE, and 4.13× for QAOA circuits compared to other simulators. The overall speedup is calculated as the arithmetic mean of normalized performance ratios across all evaluated circuits within each benchmark category.
    
    \begin{figure}[ht]
    	\centering
    	\includegraphics[width=0.8\textwidth]{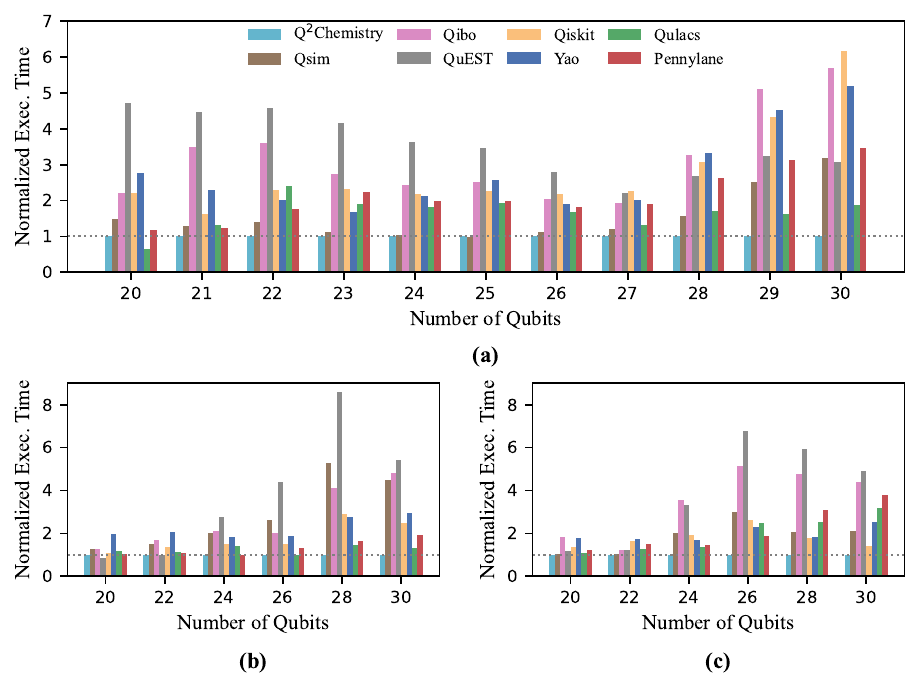}
        \caption{Normalized execution time comparison of (a) QFT, (b) VQE-HEA, and (c) QAOA circuit simulations on GPU heterogeneous platform with one A100. Results are normalized to Q$^2$Chemistry performance. }
    	\label{fig:gpu-single-benchmark}
    \end{figure}
    
\begin{figure}[ht]
    	\centering
    	\includegraphics[width=0.8\textwidth]{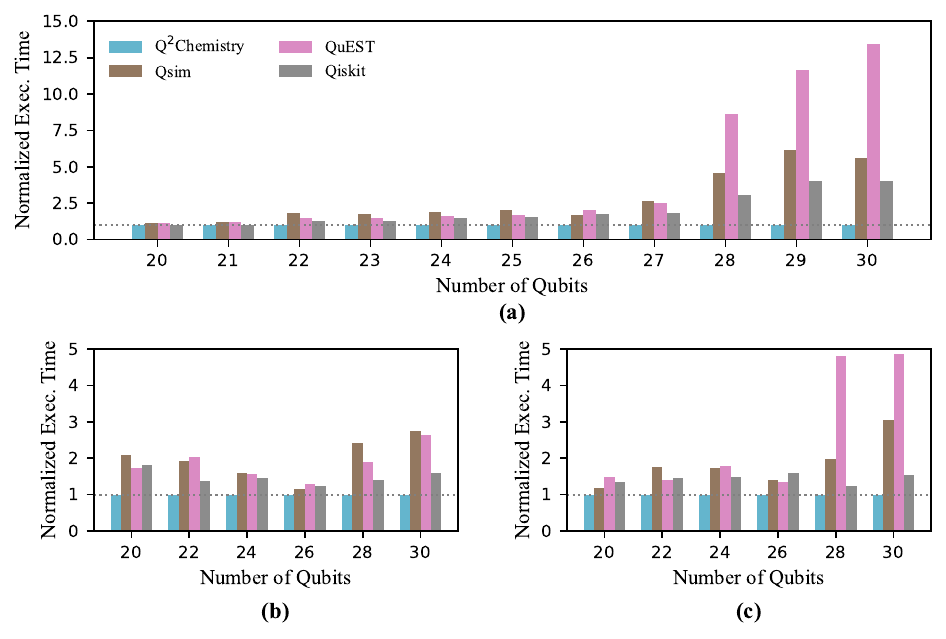}
        \caption{Normalized execution time comparison of (a) QFT, (b) VQE-HEA, and (c) QAOA circuit simulations on GPU heterogeneous platform with four A100. Results are normalized to Q$^2$Chemistry performance.}
    	\label{fig:gpu-multi-benchmark}
    \end{figure}

    In single-GPU configurations (Figure~\ref{fig:gpu-single-benchmark}), Q$^2$Chemistry extends its advantage through SMGP's two-dimensional thread blocks. For QFT circuits, it delivers up to 6x speedup over Qiskit in QFT applications, and continues to lead in most VQE-HEA and QAOA benchmarks especially with increased number of qubits. When scaling to four A100 GPUs, it achieves as large as 13.44× speedup over QuEST in 30-qubit QFT simulations and outperform other multi-GPU-enabled simulators (Qsim, Qiskit) in all test cases (Figure~\ref{fig:gpu-multi-benchmark}). This performance advantage is largely attributed to our optimized multi-GPU communication strategy, leveraging the hybrid parallel scheduling which substantially reduces inter-GPU communication overhead and synchronization delays. Unlike simulators such as Qibo, Yao, MindQuantum, and Qulacs, which do not support multi-GPU execution, Q$^2$Chemistry can efficiently harness heterogeneous multi-node GPU clusters, enabling both faster simulation and larger qubit capacity. This makes Q$^2$Chemistry particularly well-suited for high-performance quantum simulation in distributed and scalable GPU computing environments.
    
    \section{Conclusion}\label{sec5}
    In this work, we presented a comprehensive parallel optimization solution for the quantum circuit simulator in the software package Q$^2$Chemistry. Our strategy integrates three core methodologies: batch-buffered overlap processing to enable concurrent computation and communication through multi-buffering, staggered multi-gate parallelism to enhance GPU throughput through staggered execution of quantum gates, and dependency-aware gate contraction for optimizing gate fusion by analyzing control-target dependencies. Simulation bottlenecks in memory-intensive and communication-heavy scenarios are addressed, as well as optimizing the total count of sequential gate operations for a given circuit. As presented in the benchmark results, Q$^2$Chemistry delivers superior performance over state-of-the-art quantum simulators across all circuit types and hardware platforms after implementing these optimizations. These results highlight Q$^2$Chemistry's exceptional scalability and efficiency, making it well-suited for deployment on modern HPC platforms as a powerful and effective solution for both quantum chemistry and general-purpose quantum circuit simulation. Future work will focus on extending the framework to domestic accelerators such as Sunway TaihuLight processes and Hygon DCU cards, and integrating noise-aware metrics such as pre-estimated state purity and quantum Fisher information matrices to develop high-performance density-matrix-based noisy quantum simulators for algorithms on NISQ devices.
    
    \section{Acknowledgments}\label{sec6}
    The Q$^2$Chemistry software package is available on SCNet web store\cite{scnet2025} for free under Apache License 2.0. The optimized simulator backend in this research is available on Zenodo as a drop-in extension for Q$^2$Chemistry\cite{qco2025}. This work is supported by the National Natural Science Foundation of China (22393913) and the Strategic Priority Research Program of the Chinese Academy of Sciences (XDB0450000). The AI-driven experiments, simulations and model training were performed on the robotic AI-Scientist platform of Chinese Academy of Science (22393913). We also acknowledge the USTC supercomputing center for providing computational resources for this project.

    \bibliographystyle{unsrtnat}
    \bibliography{citations}

\begin{thebibliography}{58}
\providecommand{\natexlab}[1]{#1}
\providecommand{\url}[1]{\texttt{#1}}
\expandafter\ifx\csname urlstyle\endcsname\relax
  \providecommand{\doi}[1]{doi: #1}\else
  \providecommand{\doi}{doi: \begingroup \urlstyle{rm}\Url}\fi

\bibitem[Hohenberg and Kohn(1964)]{Hon64}
P.~Hohenberg and W.~Kohn.
\newblock Inhomogeneous electron gas.
\newblock \emph{Phys. Rev.}, 136:\penalty0 B864--B871, 1964.
\newblock \doi{10.1103/PhysRev.136.B864}.
\newblock URL \url{https://link.aps.org/doi/10.1103/PhysRev.136.B864}.

\bibitem[Kohn and Sham(1965)]{Kohn65}
W.~Kohn and L.~J. Sham.
\newblock Self-consistent equations including exchange and correlation effects.
\newblock \emph{Phys. Rev.}, 140:\penalty0 A1133--A1138, 1965.
\newblock \doi{10.1103/PhysRev.140.A1133}.
\newblock URL \url{https://link.aps.org/doi/10.1103/PhysRev.140.A1133}.

\bibitem[Kohn et~al.(1996)Kohn, Becke, and Parr]{Kohn96}
W.~Kohn, A.~D. Becke, and R.~G. Parr.
\newblock Density functional theory of electronic structure.
\newblock \emph{J.~Phys. Chem.}, 100:\penalty0 12974--12980, 1996.
\newblock \doi{10.1021/jp960669l}.
\newblock URL \url{https://doi.org/10.1021/jp960669l}.

\bibitem[Cohen et~al.(2012)Cohen, {Mori-Sanchez}, and Yang]{CohMorYan12}
A.~J. Cohen, P.~{Mori-Sanchez}, and W.~Yang.
\newblock Challenges for density functional theory.
\newblock \emph{Chem. Rev.}, 112:\penalty0 289--320, 2012.
\newblock \doi{10.1021/cr200107z}.
\newblock URL \url{https://doi.org/10.1021/cr200107z}.

\bibitem[Vogiatzis et~al.(2017)Vogiatzis, Ma, Olsen, Gagliardi, and
  de~Jong]{casscf24o24e}
Konstantinos~D. Vogiatzis, Dongxia Ma, Jeppe Olsen, Laura Gagliardi, and
  Wibe~A. de~Jong.
\newblock Pushing configuration-interaction to the limit: Towards massively
  parallel mcscf calculations.
\newblock \emph{The Journal of Chemical Physics}, 147\penalty0 (18):\penalty0
  184111, 11 2017.
\newblock ISSN 0021-9606.
\newblock \doi{10.1063/1.4989858}.
\newblock URL \url{https://doi.org/10.1063/1.4989858}.

\bibitem[Feynman(1982)]{Feynman1982}
Richard~P. Feynman.
\newblock Simulating physics with computers.
\newblock \emph{Int. J. Theor. Phys.}, 21\penalty0 (6):\penalty0 467--488, Jun
  1982.
\newblock ISSN 1572-9575.
\newblock \doi{10.1007/BF02650179}.
\newblock URL \url{https://doi.org/10.1007/BF02650179}.

\bibitem[Shor(1997)]{shor1997}
Peter~W. Shor.
\newblock Polynomial-time algorithms for prime factorization and discrete
  logarithms on a quantum computer.
\newblock \emph{SIAM Journal on Computing}, 26\penalty0 (5):\penalty0
  1484--1509, 1997.
\newblock \doi{10.1137/S0097539795293172}.
\newblock URL \url{https://doi.org/10.1137/S0097539795293172}.

\bibitem[Du et~al.(2010)Du, Xu, Peng, Wang, Wu, and Lu]{h2molecule2010}
Jiangfeng Du, Nanyang Xu, Xinhua Peng, Pengfei Wang, Sanfeng Wu, and Dawei Lu.
\newblock Nmr implementation of a molecular hydrogen quantum simulation with
  adiabatic state preparation.
\newblock \emph{Phys. Rev. Lett.}, 104:\penalty0 030502, Jan 2010.
\newblock \doi{10.1103/PhysRevLett.104.030502}.
\newblock URL \url{https://link.aps.org/doi/10.1103/PhysRevLett.104.030502}.

\bibitem[Arute et~al.(2019)Arute, Arya, Babbush, Bacon, Bardin, Barends,
  Biswas, Boixo, Brand{\~a}o, Buell, Burkett, Chen, Chen, Chiaro, Collins,
  Courtney, Dunsworth, Farhi, Foxen, Fowler, Gidney, Giustina, Graff, Guerin,
  Habegger, Harrigan, Hartmann, Ho, Hoffmann, Huang, Humble, Isakov, Jeffrey,
  Jiang, Kafri, Kechedzhi, Kelly, Klimov, Knysh, Korotkov, Kostritsa, Landhuis,
  Lindmark, Lucero, Lyakh, Mandr{\`a}, McClean, McEwen, Megrant, Mi,
  Michielsen, Mohseni, Mutus, Naaman, Neeley, Neill, Niu, Ostby, Petukhov,
  Platt, Quintana, Rieffel, Roushan, Rubin, Sank, Satzinger, Smelyanskiy, Sung,
  Trevithick, Vainsencher, Villalonga, White, Yao, Yeh, Zalcman, Neven, and
  Martinis]{Arute2019QuantumSU}
Frank Arute, Kunal Arya, Ryan Babbush, Dave Bacon, Joseph~C. Bardin, Rami
  Barends, Rupak Biswas, Sergio Boixo, Fernando G. S.~L. Brand{\~a}o, David~A.
  Buell, Brian Burkett, Yu~Chen, Zijun Chen, Benjamin Chiaro, Roberto Collins,
  William Courtney, Andrew Dunsworth, Edward Farhi, Brooks Foxen, Austin~G.
  Fowler, Craig Gidney, Marissa Giustina, Rob Graff, Keith Guerin, S.~Habegger,
  Matthew~P. Harrigan, Michael~J. Hartmann, Alan~K. Ho, Markus Hoffmann, Trent
  Huang, T.~Humble, Sergei~V. Isakov, Evan Jeffrey, Zhang Jiang, Dvir Kafri,
  Kostyantyn Kechedzhi, Julian Kelly, Paul~V. Klimov, Sergey Knysh,
  Alexander~N. Korotkov, Fedor Kostritsa, David Landhuis, Mike Lindmark, Erik
  Lucero, Dmitry~I. Lyakh, Salvatore Mandr{\`a}, Jarrod~R. McClean, Matthew~J.
  McEwen, Anthony Megrant, Xiao Mi, Kristel Michielsen, Masoud Mohseni, Josh
  Mutus, Ofer Naaman, Matthew Neeley, Charles~J. Neill, Murphy~Yuezhen Niu,
  Eric~P. Ostby, Andre Petukhov, John~C. Platt, Chris Quintana, Eleanor~Gilbert
  Rieffel, Pedram Roushan, Nicholas~C. Rubin, Daniel~Thomas Sank, Kevin~J.
  Satzinger, Vadim~N. Smelyanskiy, Kevin~J. Sung, Matthew~D. Trevithick, Amit
  Vainsencher, Benjamin Villalonga, Theodore White, Z.~Yao, Ping Yeh, Adam
  Zalcman, Hartmut Neven, and John~M. Martinis.
\newblock Quantum supremacy using a programmable superconducting processor.
\newblock \emph{Nature}, 574:\penalty0 505--510, 2019.
\newblock \doi{10.1038/s41586-019-1666-5}.

\bibitem[Tilly et~al.(2021)Tilly, Chen, Cao, et~al.]{Tilly_VQE_2021}
Jules Tilly, Hongxiang Chen, Shuxiang Cao, et~al.
\newblock The variational quantum eigensolver: a review of methods and best
  practices, 2021.
\newblock URL \url{https://arxiv.org/abs/2111.05176}.
\newblock \url{{https://arxiv.org/abs/2111.05176}}, Accessed August 1, 2022.

\bibitem[Cerezo et~al.(2021)Cerezo, Arrasmith, Babbush, et~al.]{Cerezo2021}
M.~Cerezo, Andrew Arrasmith, Ryan Babbush, et~al.
\newblock Variational quantum algorithms.
\newblock \emph{Nat. Rev. Phys.}, 3\penalty0 (9):\penalty0 625--644, Sep 2021.
\newblock ISSN 2522-5820.
\newblock \doi{10.1038/s42254-021-00348-9}.
\newblock URL \url{https://doi.org/10.1038/s42254-021-00348-9}.

\bibitem[Magann et~al.(2021)Magann, Arenz, Grace, et~al.]{Magann_VQA_2021}
Alicia~B. Magann, Christian Arenz, Matthew~D. Grace, et~al.
\newblock From pulses to circuits and back again: A quantum optimal control
  perspective on variational quantum algorithms.
\newblock \emph{Phys. Rev. X. Quantum}, 2:\penalty0 010101, Jan 2021.
\newblock \doi{10.1103/PRXQuantum.2.010101}.
\newblock URL \url{https://link.aps.org/doi/10.1103/PRXQuantum.2.010101}.

\bibitem[Fedorov et~al.(2022)Fedorov, Peng, Govind,
  et~al.]{Fedorov_VQERev_2022}
Dmitry~A. Fedorov, Bo~Peng, Niranjan Govind, et~al.
\newblock Vqe method: a short survey and recent developments.
\newblock \emph{Materials Theory}, 6\penalty0 (1):\penalty0 2, Jan 2022.
\newblock ISSN 2509-8012.
\newblock \doi{10.1186/s41313-021-00032-6}.
\newblock URL \url{https://doi.org/10.1186/s41313-021-00032-6}.

\bibitem[Bravyi and Kitaev(2002)]{BraKit02}
S.~B. Bravyi and A.~Y. Kitaev.
\newblock Fermionic quantum computation.
\newblock \emph{Ann. Phys.}, 298:\penalty0 210--226, 2002.
\newblock \doi{https://doi.org/10.1006/aphy.2002.6254}.
\newblock URL
  \url{https://www.sciencedirect.com/science/article/pii/S0003491602962548}.

\bibitem[McArdle et~al.(2020)McArdle, Endo, Aspuru-Guzik, et~al.]{McAEndAsp20}
Sam McArdle, Suguru Endo, Al\'an Aspuru-Guzik, et~al.
\newblock Quantum computational chemistry.
\newblock \emph{Rev. Mod. Phys.}, 92:\penalty0 015003, 2020.
\newblock \doi{10.1103/RevModPhys.92.015003}.
\newblock URL \url{https://link.aps.org/doi/10.1103/RevModPhys.92.015003}.

\bibitem[Cao et~al.(2019)Cao, Romero, Olson, et~al.]{CaoRomOls19}
Yudong Cao, Jonathan Romero, Jonathan~P. Olson, et~al.
\newblock Quantum chemistry in the age of quantum computing.
\newblock \emph{Chem. Rev.}, 119:\penalty0 10856--10915, 2019.
\newblock \doi{10.1021/acs.chemrev.8b00803}.
\newblock URL \url{https://doi.org/10.1021/acs.chemrev.8b00803}.

\bibitem[Preskill(2018)]{Preskill_2018}
John Preskill.
\newblock Quantum computing in the nisq era and beyond.
\newblock \emph{Quantum}, 2:\penalty0 79, August 2018.
\newblock ISSN 2521-327X.
\newblock \doi{10.22331/q-2018-08-06-79}.

\bibitem[Georgescu et~al.(2014)Georgescu, Ashhab, and Nori]{GeoAshNor14}
I.~M. Georgescu, S.~Ashhab, and Franco Nori.
\newblock Quantum simulation.
\newblock \emph{Rev. Mod. Phys.}, 86:\penalty0 153--185, 2014.
\newblock \doi{10.1103/RevModPhys.86.153}.
\newblock URL \url{https://link.aps.org/doi/10.1103/RevModPhys.86.153}.

\bibitem[Aspuru-Guzik et~al.(2005)Aspuru-Guzik, Dutoi, Love,
  et~al.]{AspDutLov05}
A.~Aspuru-Guzik, A.~D. Dutoi, P.~J. Love, et~al.
\newblock Simulated quantum computation of molecular energies.
\newblock \emph{Science}, 309:\penalty0 1704--1707, 2005.
\newblock \doi{10.1126/science.1113479}.
\newblock URL \url{https://science.sciencemag.org/content/309/5741/1704}.

\bibitem[Wang et~al.(2008)Wang, Kais, Aspuru-Guzik, et~al.]{Wang08}
Hefeng Wang, Sabre Kais, Al{\'a}n Aspuru-Guzik, et~al.
\newblock Quantum algorithm for obtaining the energy spectrum of molecular
  systems.
\newblock \emph{Phys. Chem. Chem. Phys.}, 10:\penalty0 5388--5393, 2008.
\newblock \doi{10.1039/B804804E}.
\newblock URL \url{http://dx.doi.org/10.1039/B804804E}.

\bibitem[Peruzzo et~al.(2014)Peruzzo, McClean, Shadbolt, et~al.]{PerMcCSha14}
A.~Peruzzo, J.~McClean, P.~Shadbolt, et~al.
\newblock A variational eigenvalue solver on a photonic quantum processor.
\newblock \emph{Nat. Commun.}, 5:\penalty0 4213, 2014.
\newblock \doi{10.1038/ncomms5213}.
\newblock URL \url{https://doi.org/10.1038/ncomms5213}.

\bibitem[Hempel et~al.(2018)Hempel, Maier, Romero, et~al.]{HemMaiRom18}
C.~Hempel, C.~Maier, J.~Romero, et~al.
\newblock Quantum chemistry calculations on a trapped-ion quantum simulator.
\newblock \emph{Phys. Rev. X}, 8:\penalty0 031022, 2018.
\newblock \doi{10.1103/PhysRevX.8.031022}.
\newblock URL \url{https://link.aps.org/doi/10.1103/PhysRevX.8.031022}.

\bibitem[Nam et~al.(2020)Nam, Chen, Pisenti, et~al.]{NamChen20}
Yunseong Nam, Jwo~Sy Chen, Neal~C. Pisenti, et~al.
\newblock Ground-state energy estimation of the water molecule on a trapped-ion
  quantum computer.
\newblock \emph{Npj Quantum Inf.}, 6:\penalty0 33, 2020.
\newblock \doi{10.1038/s41534-020-0259-3}.
\newblock URL \url{https://doi.org/10.1038/s41534-020-0259-3}.

\bibitem[Shen et~al.(2017)Shen, Zhang, Zhang, et~al.]{SheZhaZha17}
Y.~Shen, X.~Zhang, S.~Zhang, et~al.
\newblock Quantum implementation of the unitary coupled cluster for simulating
  molecular electronic structure.
\newblock \emph{Phys. Rev. A: At., Mol., Opt. Phys.}, 95:\penalty0 020501,
  2017.
\newblock \doi{10.1103/PhysRevA.95.020501}.
\newblock URL \url{https://link.aps.org/doi/10.1103/PhysRevA.95.020501}.

\bibitem[O'~Malley et~al.(2016)O'~Malley, Babbush, Kivlichan,
  et~al.]{MalBabKiv16}
P.~J.~J. O'~Malley, R.~Babbush, I.~D. Kivlichan, et~al.
\newblock Scalable quantum simulation of molecular energies.
\newblock \emph{Phys. Rev. X}, 6:\penalty0 031007, 2016.
\newblock \doi{10.1103/PhysRevX.6.031007}.
\newblock URL \url{https://link.aps.org/doi/10.1103/PhysRevX.6.031007}.

\bibitem[Kandala et~al.(2017)Kandala, Mezzacapo, Temme, et~al.]{Kandala2017}
Abhinav Kandala, Antonio Mezzacapo, Kristan Temme, et~al.
\newblock Hardware-efficient variational quantum eigensolver for small
  molecules and quantum magnets.
\newblock \emph{Nature}, 549\penalty0 (7671):\penalty0 242--246, Sep 2017.
\newblock ISSN 1476-4687.
\newblock \doi{10.1038/nature23879}.
\newblock URL \url{https://doi.org/10.1038/nature23879}.

\bibitem[Colless et~al.(2018)Colless, Ramasesh, Dahlen, et~al.]{ColRamDah18}
J.~I. Colless, V.~V. Ramasesh, D.~Dahlen, et~al.
\newblock Computation of molecular spectra on a quantum processor with an
  error-resilient algorithm.
\newblock \emph{Phys. Rev. X}, 8:\penalty0 011021, 2018.
\newblock \doi{10.1103/PhysRevX.8.011021}.
\newblock URL \url{https://link.aps.org/doi/10.1103/PhysRevX.8.011021}.

\bibitem[McClean et~al.(2016)McClean, Romero, Babbush, et~al.]{McCRomBab16}
J.~R. McClean, J.~Romero, R.~Babbush, et~al.
\newblock The theory of variational hybrid quantum-classical algorithms.
\newblock \emph{New J. Phys.}, 18:\penalty0 023023, 2016.
\newblock \doi{10.1088/1367-2630/18/2/023023}.
\newblock URL \url{https://doi.org/10.1088/1367-2630/18/2/023023}.

\bibitem[Lanyon et~al.(2010)Lanyon, Whitfield, Gillett, et~al.]{LanWhi10}
B.~P. Lanyon, J.~D. Whitfield, G.~G. Gillett, et~al.
\newblock Towards quantum chemistry on a quantum computer.
\newblock \emph{Nat. Chem.}, 2:\penalty0 106--111, 2010.
\newblock \doi{10.1038/nchem.483}.
\newblock URL \url{https://doi.org/10.1038/nchem.483}.

\bibitem[Romero et~al.(2018)Romero, Babbush, McClean, et~al.]{RomBabMcC18}
J.~Romero, R.~Babbush, J.~R. McClean, et~al.
\newblock Strategies for quantum computing molecular energies using the unitary
  coupled cluster ansatz.
\newblock \emph{Quantum Sci. Technol.}, 4:\penalty0 014008, 2018.
\newblock \doi{10.1088/2058-9565/aad3e4}.
\newblock URL \url{https://doi.org/10.1088/2058-9565/aad3e4}.

\bibitem[Higgott et~al.(2019)Higgott, Wang, and Brierley]{vqe-excited-vqd}
Oscar Higgott, Daochen Wang, and Stephen Brierley.
\newblock Variational {Q}uantum {C}omputation of {E}xcited {S}tates.
\newblock \emph{{Quantum}}, 3:\penalty0 156, 2019.
\newblock \doi{10.22331/q-2019-07-01-156}.
\newblock URL \url{https://doi.org/10.22331/q-2019-07-01-156}.

\bibitem[McClean et~al.(2017)McClean, Kimchi-Schwartz, Carter,
  et~al.]{Mcclean17qse}
Jarrod~R. McClean, Mollie~E. Kimchi-Schwartz, Jonathan Carter, et~al.
\newblock Hybrid quantum-classical hierarchy for mitigation of decoherence and
  determination of excited states.
\newblock \emph{Phys. Rev. A}, 95:\penalty0 042308, Apr 2017.
\newblock \doi{10.1103/PhysRevA.95.042308}.
\newblock URL \url{https://link.aps.org/doi/10.1103/PhysRevA.95.042308}.

\bibitem[Yung et~al.(2014)Yung, Casanova, Mezzacapo, et~al.]{YungCas2014}
Man~Hong Yung, J.~Casanova, A.~Mezzacapo, et~al.
\newblock From transistor to trapped-ion computers for quantum chemistry.
\newblock \emph{Sci. Rep.}, 4\penalty0 (1):\penalty0 3589, Jan 2014.
\newblock \doi{10.1038/srep03589}.
\newblock URL \url{https://doi.org/10.1038/srep03589}.

\bibitem[Farhi et~al.(2014)Farhi, Goldstone, and Gutmann]{farhi2014}
Edward Farhi, Jeffrey Goldstone, and Sam Gutmann.
\newblock A quantum approximate optimization algorithm, 2014.
\newblock URL \url{https://arxiv.org/abs/1411.4028}.

\bibitem[Bartlett et~al.(1989)Bartlett, Kucharski, and Noga]{BarKucNog89}
R.~J. Bartlett, S.~A. Kucharski, and J.~Noga.
\newblock Alternative coupled-cluster ans\"atze ii. the unitary coupled-cluster
  method.
\newblock \emph{Chem. Phys. Lett.}, 155:\penalty0 133--140, 1989.
\newblock \doi{https://doi.org/10.1016/S0009-2614(89)87372-5}.
\newblock URL
  \url{https://www.sciencedirect.com/science/article/pii/S0009261489873725}.

\bibitem[Taube and Bartlett(2006)]{TauBar06}
A.~G. Taube and R.~J. Bartlett.
\newblock New perspectives on unitary coupled-cluster theory.
\newblock \emph{Int. J. Quantum Chem.}, 106:\penalty0 3393--3401, 2006.
\newblock \doi{https://doi.org/10.1002/qua.21198}.
\newblock URL \url{https://onlinelibrary.wiley.com/doi/abs/10.1002/qua.21198}.

\bibitem[Guo et~al.(2023)Guo, Fan, Xu, and Shang]{Guo2023differentiable}
Chu Guo, Yi~Fan, Zhiqian Xu, and Honghui Shang.
\newblock Differentiable matrix product states for simulating variational
  quantum computational chemistry.
\newblock \emph{{Quantum}}, 7:\penalty0 1192, December 2023.
\newblock ISSN 2521-327X.
\newblock \doi{10.22331/q-2023-12-04-1192}.
\newblock URL \url{https://doi.org/10.22331/q-2023-12-04-1192}.

\bibitem[Vidal(2007)]{Vidal2007MPS}
G.~Vidal.
\newblock Classical simulation of infinite-size quantum lattice systems in one
  spatial dimension.
\newblock \emph{Phys. Rev. Lett.}, 98:\penalty0 070201, Feb 2007.
\newblock \doi{10.1103/PhysRevLett.98.070201}.
\newblock URL \url{https://link.aps.org/doi/10.1103/PhysRevLett.98.070201}.

\bibitem[Hastings(2009)]{Hastings2009MPS}
M.~B. Hastings.
\newblock Light-cone matrix product.
\newblock \emph{Journal of Mathematical Physics}, 50\penalty0 (9):\penalty0
  095207, 06 2009.
\newblock ISSN 0022-2488.
\newblock \doi{10.1063/1.3149556}.
\newblock URL \url{https://doi.org/10.1063/1.3149556}.

\bibitem[Efthymiou et~al.(2021)Efthymiou, Ramos-Calderer, Bravo-Prieto,
  Pérez-Salinas, García-Martín, Garcia-Saez, Latorre, and Carrazza]{Qibo}
Stavros Efthymiou, Sergi Ramos-Calderer, Carlos Bravo-Prieto, Adrián
  Pérez-Salinas, Diego García-Martín, Artur Garcia-Saez, José~Ignacio
  Latorre, and Stefano Carrazza.
\newblock Qibo: a framework for quantum simulation with hardware acceleration.
\newblock \emph{Quantum Science and Technology}, 7\penalty0 (1):\penalty0
  015018, December 2021.
\newblock ISSN 2058-9565.
\newblock \doi{10.1088/2058-9565/ac39f5}.
\newblock URL \url{http://dx.doi.org/10.1088/2058-9565/ac39f5}.

\bibitem[Avila et~al.(2014)Avila, Maron, Reiser, Pilla, and
  Yamin]{Avila2014GPUaware}
Anderson Avila, Adriano Maron, Renata Reiser, Maur{\'i}cio Pilla, and Adenauer
  Yamin.
\newblock Gpu-aware distributed quantum simulation.
\newblock In \emph{Proceedings of the 29th Annual ACM Symposium on Applied
  Computing}, pages 893--900, New York, NY, USA, March 2014. ACM.
\newblock \doi{10.1145/2554850.2554892}.

\bibitem[Doi et~al.(2019)Doi, Takahashi, Putra, Imamichi, and
  Horii]{2019QuantumCS}
Jun Doi, Hitomi Takahashi, Raymond~H. Putra, Takashi Imamichi, and Hiroshi
  Horii.
\newblock Quantum computing simulator on a heterogenous hpc system.
\newblock \emph{Proceedings of the 16th ACM International Conference on
  Computing Frontiers}, 2019.
\newblock \doi{10.1145/3310273.3323053}.
\newblock URL \url{https://api.semanticscholar.org/CorpusID:153313935}.

\bibitem[Gutierrez et~al.(2007)Gutierrez, Romero, Trenas, and
  Zapata]{Gutierrez}
Eladio Gutierrez, Sergio Romero, Maria~A. Trenas, and Emilio~L. Zapata.
\newblock Simulation of quantum gates on a novel gpu architecture.
\newblock In \emph{Proceedings of the 7th WSEAS International Conference on
  Systems Theory and Scientific Computation}, pages 121--126, Stevens Point,
  Wisconsin, USA, 2007. World Scientific and Engineering Academy and Society
  (WSEAS).
\newblock ISBN 9789608457980.
\newblock \doi{10.1103/physreva.94.062304}.

\bibitem[Guti{\'e}rrez et~al.(2008)Guti{\'e}rrez, Romero, Trenas, and
  Zapata]{Gutirrez2008ParallelQC}
Eladio Guti{\'e}rrez, Sergio Romero, Mar{\'i}a~A. Trenas, and Emilio~L. Zapata.
\newblock Parallel quantum computer simulation on the cuda architecture.
\newblock In \emph{International Conference on Conceptual Structures}, 2008.
\newblock \doi{10.1007/978-3-540-69384-0_75}.
\newblock URL \url{https://api.semanticscholar.org/CorpusID:37760018}.

\bibitem[H\"{a}ner and Steiger(2017)]{0.5petabyte}
Thomas H\"{a}ner and Damian~S. Steiger.
\newblock 0.5 petabyte simulation of a 45-qubit quantum circuit.
\newblock In \emph{Proceedings of the International Conference for High
  Performance Computing, Networking, Storage and Analysis}, SC '17, New York,
  NY, USA, 2017. Association for Computing Machinery.
\newblock ISBN 9781450351140.
\newblock \doi{10.1145/3126908.3126947}.
\newblock URL \url{https://doi.org/10.1145/3126908.3126947}.

\bibitem[De~Raedt et~al.(2007)De~Raedt, Michielsen, De~Raedt, Trieu, Arnold,
  Richter, Lippert, Watanabe, and Ito]{massive-parallel}
K.~De~Raedt, K.~Michielsen, H.~De~Raedt, B.~Trieu, G.~Arnold, M.~Richter, Th.
  Lippert, H.~Watanabe, and N.~Ito.
\newblock Massively parallel quantum computer simulator.
\newblock \emph{Computer Physics Communications}, 176\penalty0 (2):\penalty0
  121–136, 2007.
\newblock ISSN 0010-4655.
\newblock \doi{10.1016/j.cpc.2006.08.007}.
\newblock URL \url{http://dx.doi.org/10.1016/j.cpc.2006.08.007}.

\bibitem[Smelyanskiy et~al.(2016)Smelyanskiy, Sawaya, and
  Aspuru-Guzik]{qhipster}
Mikhail Smelyanskiy, Nicolas P.~D. Sawaya, and Alán Aspuru-Guzik.
\newblock qhipster: The quantum high performance software testing environment,
  2016.

\bibitem[Zulehner and Wille(2018)]{zulehner2018}
Alwin Zulehner and Robert Wille.
\newblock Advanced simulation of quantum computations, 2018.
\newblock URL \url{https://arxiv.org/abs/1707.00865}.

\bibitem[Wu et~al.(2019)Wu, Di, Dasgupta, Cappello, Finkel, Alexeev, and
  Chong]{Wu_2019}
Xin-Chuan Wu, Sheng Di, Emma~Maitreyee Dasgupta, Franck Cappello, Hal Finkel,
  Yuri Alexeev, and Frederic~T. Chong.
\newblock Full-state quantum circuit simulation by using data compression.
\newblock In \emph{Proceedings of the International Conference for High
  Performance Computing, Networking, Storage and Analysis}, SC '19, New York,
  NY, USA, 2019. Association for Computing Machinery.
\newblock ISBN 9781450362290.
\newblock \doi{10.1145/3295500.3356155}.
\newblock URL \url{https://doi.org/10.1145/3295500.3356155}.

\bibitem[Aleksandrowicz et~al.(2019)Aleksandrowicz, Alexander, Barkoutsos,
  Bello, Ben-Haim, Bucher, Cabrera-Hern{\'a}ndez, Carballo-Franquis, Chen,
  Chen, Chow, C{\'o}rcoles-Gonzales, Cross, Cross, Cruz-Benito,
  et~al.]{qiskit-package}
Gadi Aleksandrowicz, Thomas Alexander, Panagiotis Barkoutsos, Luciano Bello,
  Yael Ben-Haim, David Bucher, Francisco~Jose Cabrera-Hern{\'a}ndez, Jorge
  Carballo-Franquis, Adrian Chen, Chun-Fu Chen, Jerry~M. Chow, Antonio~D.
  C{\'o}rcoles-Gonzales, Abigail~J. Cross, Andrew Cross, Juan Cruz-Benito,
  et~al.
\newblock Qiskit: An open-source framework for quantum computing, January 2019.
\newblock Version 0.7.2.

\bibitem[Zhang et~al.(2015)Zhang, Yuan, and Lu]{Zhang-2015}
Pei Zhang, Jiabin Yuan, and Xiangwen Lu.
\newblock Quantum computer simulation on multi-gpu incorporating data locality.
\newblock In \emph{Algorithms and Architectures for Parallel Processing (ICA3PP
  2015)}, volume 9528, pages 241--256, 11 2015.
\newblock ISBN 978-3-319-27118-7.
\newblock \doi{10.1007/978-3-319-27119-4_17}.

\bibitem[{Cirq Developers}(2022)]{cirq}
{Cirq Developers}.
\newblock {Cirq}, April 2022.
\newblock URL \url{https://zenodo.org/doi/10.5281/zenodo.4062499}.
\newblock Version v0.14.1; accessed 2022-08-01.

\bibitem[Suzuki et~al.(2021)Suzuki, Kawase, Masumura, Hiraga, Nakadai, Chen,
  Nakanishi, Mitarai, Imai, Tamiya, Yamamoto, Yan, Kawakubo, Nakagawa, Ibe,
  Zhang, Yamashita, Yoshimura, Hayashi, and Fujii]{qulacs}
Yasunari Suzuki, Yoshiaki Kawase, Yuya Masumura, Yuria Hiraga, Masahiro
  Nakadai, Jiabao Chen, Ken~M. Nakanishi, Kosuke Mitarai, Ryosuke Imai, Shiro
  Tamiya, Takahiro Yamamoto, Tennin Yan, Toru Kawakubo, Yuya~O. Nakagawa, Yohei
  Ibe, Youyuan Zhang, Hirotsugu Yamashita, Hikaru Yoshimura, Akihiro Hayashi,
  and Keisuke Fujii.
\newblock Qulacs: a fast and versatile quantum circuit simulator for research
  purpose.
\newblock \emph{Quantum}, 5:\penalty0 559, October 2021.
\newblock \doi{10.22331/q-2021-10-06-559}.

\bibitem[Jones et~al.(2019)Jones, Brown, Bush, and Benjamin]{Jones_QUEST_2019}
Tyson Jones, Anna Brown, Ian Bush, and Simon~C. Benjamin.
\newblock {QuEST} and high performance simulation of quantum computers.
\newblock \emph{Scientific Reports}, 9\penalty0 (1):\penalty0 10736, July 2019.
\newblock ISSN 2045-2322.
\newblock \doi{10.1038/s41598-019-47174-9}.

\bibitem[Fan et~al.(2022)Fan, Liu, Zeng, Xu, Shang, Li, and Yang]{q2chem}
Yi~Fan, Jie Liu, Xiongzhi Zeng, Zhiqian Xu, Honghui Shang, Zhenyu Li, and
  Jinlong Yang.
\newblock Q$^2$chemistry: A quantum computation platform for quantum chemistry.
\newblock \emph{JUSTC}, 52\penalty0 (12):\penalty0 2, 2022.
\newblock \doi{10.52396/JUSTC-2022-0118}.

\bibitem[Hiroshi et~al.(2021)Hiroshi, Jun, Hiroshi, and Jun]{gate_fusion}
Horii Hiroshi, Doi Jun, Horii Hiroshi, and Doi Jun.
\newblock Optimization of quantum computing simulation with gate fusion.
\newblock \emph{Information Processing Society of Japan}, Mar 2021.
\newblock ISSN 0167-9260.
\newblock \doi{https://doi.org/10.1016/j.vlsi.2019.10.004}.
\newblock URL
  \url{https://www.sciencedirect.com/science/article/pii/S0167926019302755}.

\bibitem[{SCNet}(2025)]{scnet2025}
{SCNet}.
\newblock {SCNet}: Supercomputing network.
\newblock
  \url{https://www.scnet.cn/ui/mall/detail/goods?type=software&shopId=1788135137565712385&common1=APP_SOFTWARE&id=1834155267946004482},
  2025.
\newblock Accessed: 2025-08-26, Software available under Apache License 2.0.

\bibitem[Zhong(2025)]{qco2025}
G.~Zhong.
\newblock Quantumsimulatoroptimizer, 2025.
\newblock URL \url{https://doi.org/10.5281/zenodo.17079328}.
\newblock Published September 8, 2025 | Version v1.

\end{thebibliography}
\end{document}